\documentclass{svmult} 

\pdfoutput=1 

\usepackage{newtxtext}       
\usepackage[varvw]{newtxmath}   

\usepackage[utf8]{inputenc} 

\usepackage[backend=biber, sorting=none]{biblatex}
\usepackage{amsmath}
\usepackage{amsfonts}
\usepackage{physics}

\usepackage{graphicx} 
\usepackage{float} 

\usepackage{array} 

\usepackage{hyperref} 
\hypersetup{
    colorlinks=true,
    linkcolor=red,
    citecolor=red,      
    urlcolor=black}

\DeclareMathOperator*{\argmax}{argmax}

\newcommand{\beql}[1]{\begin{equation}\label{#1}} 
\newcommand{\eeq}{\end{equation}}

\begin{document}

\title*{Semiclassical and Continuum Limits of Four-Dimensional CDT}
\author{Jakub Gizbert-Studnicki}

\institute{Jakub Gizbert-Studnicki \at Institute of Theoretical Physics, Jagiellonian University \\
 ul. prof. S. \L ojasiewicza 11, 30-348 Krak\'ow, Poland \\ \email{jakub.gizbert-studnicki@uj.edu.pl}}

\maketitle

\abstract {This Chapter discusses the infrared and the (perspective) ultraviolet limits of four-dimensional Causal Dynamical Triangulations (CDT). CDT is a  non-perturabtive and background-independent approach to quantization of Einstein's gravity,  based on  a lattice regularization of the quantum-gravitational path integral.
It is shown that inside the, so-called, de Sitter phase  one recovers a four-dimensional  semiclassical universe, where  behaviour of the scale factor is consistent with classical solutions of Einstein's field equations and fluctuations of the scale factor are very accurately  described by  a simple minisuperspace effective action.
It is argued that some of the phase transitions bordering the semiclassical  phase are higher-order transitions, which opens a possibility of defining a continuum limit of CDT. Finally, it is  discussed how to define the renormalization group flow in CDT and then how to search for a UV fixed point  in which, in the spirit of asymptotic safety, quantum theory of gravity could become non-perturbatively renormalizable and thus valid up to arbitrarily short distance scales. 
}

\hspace{1cm}


\newpage

\section{Introduction}\label{Sec:Intro}

Three out of four fundamental interactions  are very accurately  described by a quantum field theory (QFT), the infamous exception being the  gravitational interaction. Therefore  a quantum description of gravity remains   an important goal of theoretical physics. Despite many  different  approaches described in this Handbook, yet to date there is still no fully consistent theory of quantum gravity. One of the problems one encounters when trying to apply QFT framework to quantization of General Relativity, which is currently our best description of the classical gravitational interaction, is that such an approach is  perturbatively non-renormalizable.  It means, that the na\"ive application of the perturbation theory would result in infinitely many counterterms and related coupling constants appearing in the theory in the ultraviolet (UV) limit, that cannot be eliminated via renormalization thus yielding the theory to be unpredictive.
 However, as suggested in Weinberg’s seminal work \cite{Weinberg}, the definition of renormalizability might be generalised to the non-perturbative regime as described by the asymptotic safety conjecture. 
Asymptotic safety\footnote{For more details check the part of the Handbook dedicated to the asymptotically safe quantum gravity and for the links of asymptotic safety with CDT refer to \cite{chapterRG}.} predicts that the renormalization group (RG) flow of the bare gravitational couplings leads to the non-Gaussian UV
fixed point(s), in which the quantum  theory of gravity becomes scale-invariant (which by construction guarantees a constant limit, making
 the theory predictive), but also in the UV limit the dimensionless gravitational couplings do not need to be small (thus invalidating the na\"ive use of perturbation theory). 
The actual calculations using the functional renormalization group make approximations which can be difficult to control, and it is therefore important to verify the results by independent methods. 
This motivates a complementary lattice QFT approach which can potentially be used for testing the asymptotic safety scenario. 
In the lattice formulation the 
UV fixed point should appear as a  higher-order critical point. 
The reason is the following: the divergent correlation length, characteristic for a higher-order phase transition,  should allow  taking the lattice spacing  to zero while simultaneously keeping physical observables fixed; as the lattice spacing $a$ plays a role of a UV regulator (providing the UV cutoff of the QFT of order $1/a$), the cutoff will be removed in the continuum limit, which hopefully should correspond to the UV fixed point of quantum gravity. At the same time, the lattice QFT in question  should be able to correctly reconstruct the infrared (IR) limit of gravity, hopefully consistent with  classical Einstein's field equations. In principle one should also be able to  define how to renormalize the dimensionless bare lattice coupling constants when taking the lattice spacing to zero (or alternatively the UV cutoff to infinity).  The RG flow of the couplings with decreasing lattice spacing should define some path(s) in the lattice coupling constant space  leading from the  IR semiclassical limit to the UV continuum limit.

\begin{figure}[ht!]
\centering
    {\includegraphics[width=0.45\textwidth]{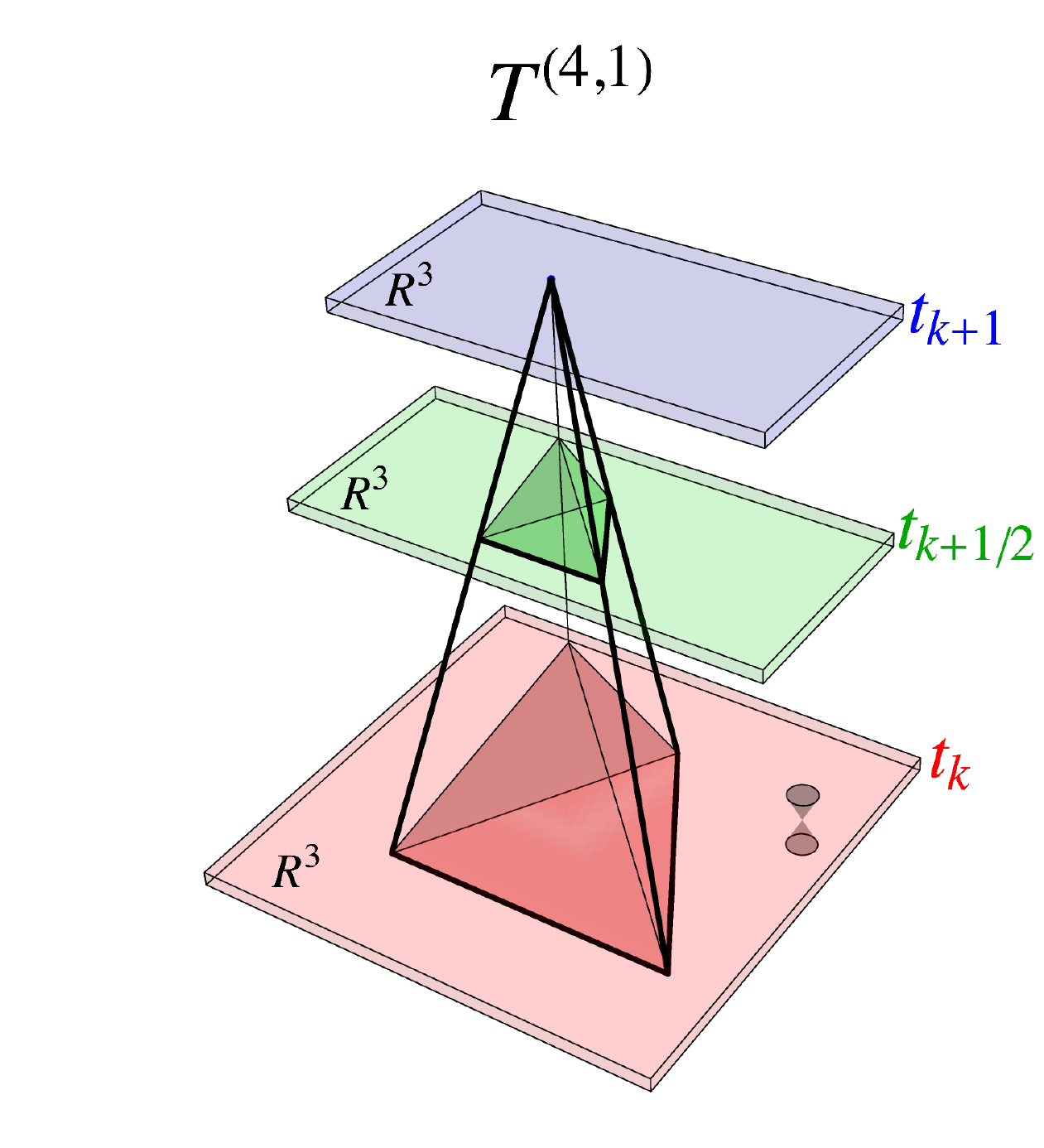}}
    {\includegraphics[width=0.45\textwidth]{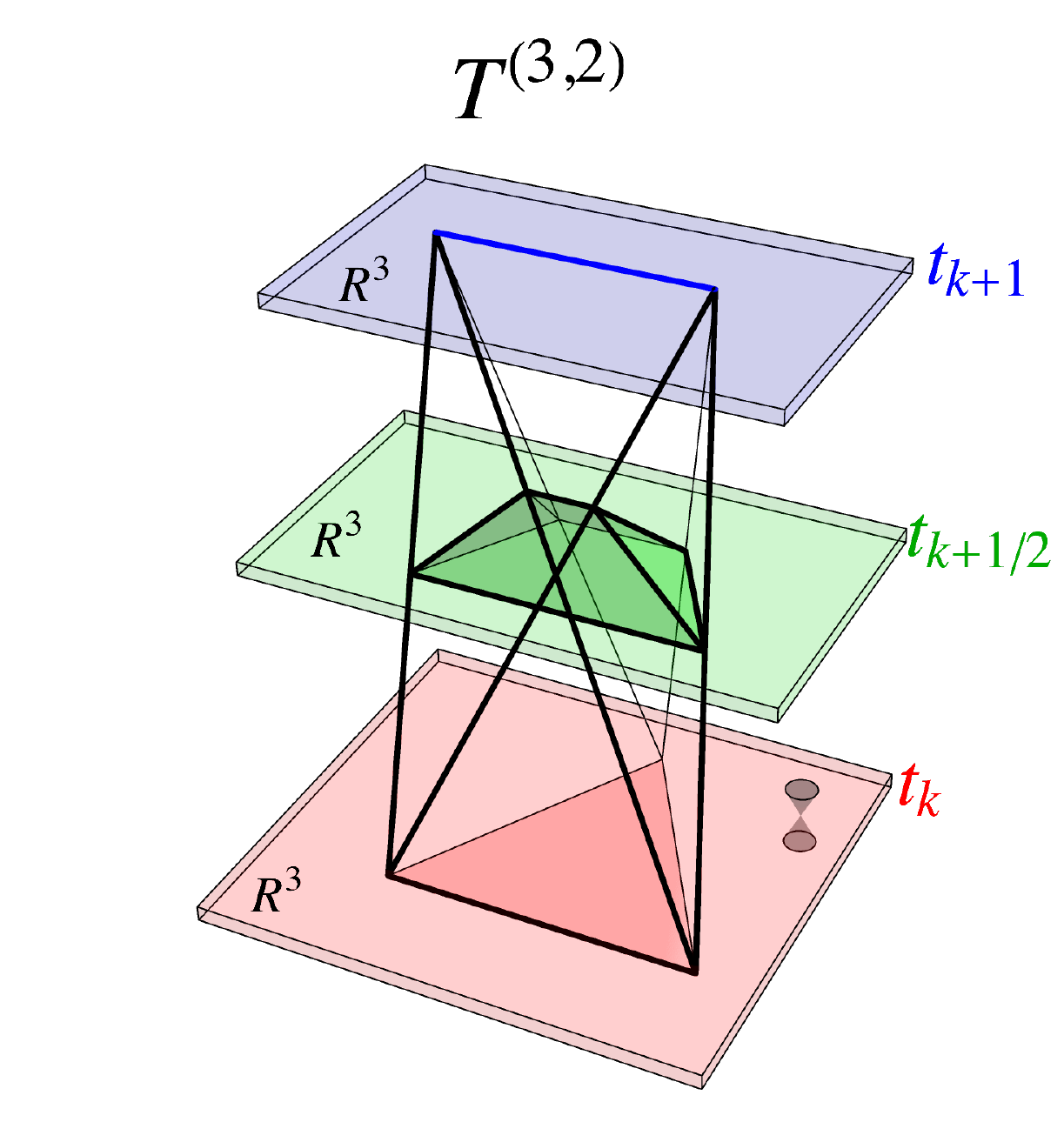}}
\caption{The building blocks of the four-dimensional CDT triangulation: the $T^{(4,1)}$ simplex (left panel) and  the $T^{(3,2)}$ simplex (right panel). The other two types, $T^{(1,4)}$ and $T^{(2,3)}$,
are time-mirrored versions of them.}
\label{fig:simplices}
\end{figure}

As will be argued below, one of the most successful lattice formulations of quantum gravity is that of Causal Dynamical Triangulations (CDT)\footnote{For  more details about the CDT formulation see \cite{chapterJan} and review articles  \cite{oldReview, RenateReview,torusReview}.}, at least in the sense that it seems to admit a well defined semiclassical description and, at the same time,  it has a rich phase structure, where some of the phase transitions might be higher-order transitions, and one  can also define how to renormalize the bare couplings of CDT when the lattice spacing is changed; an open question remains if the flow of couplings really  leads from the IR to the UV fixed point of quantum gravity. 
The  approach  
is based on the lattice regularization of the  quantum-gravitational path integral over metric d.o.f. (with Lorentzian signature)
\beql{QGPI}
    {Z}_{QG}(G, \Lambda) = \int {\cal D}[g_{\mu\nu}]e^{iS_{EH}[G,\Lambda;\, g_{\mu\nu}]}  ,
\end{equation}
where ${\cal D}[g_{\mu\nu}]$ is a measure term, which enables one to integrate over geometries, i.e., diffeomorphism-invariant equivalence classes of metrics $g_{\mu \nu}$, and $S_{EH}$ is the Einstein-Hilbert action, dependent on the Newton's constant $G$ and the cosmological constant $\Lambda$. 
In  the four-dimensional CDT the  continuous geometries are approximated by lattices (called {\it triangulations}) constructed from  four-dimensional simplicial building blocks with fixed edge lengths (called {\it 4-simplices}), fulfilling some additional topological constraints. The interior of each 4-simplex is isomorphic to a (part of) flat Minkowski spacetime and the geometry is encoded in the way the building blocks are ”glued” together.
This idea originates from earlier methods of  Euclidean Dynamical Triangulations (EDT) \cite{chapterEDT}, which assumed Euclidean (local SO(4) symmetry) instead of Lorentzian (local SO(3;1) symmetry) spacetime geometry. 
In EDT all simplices were equilateral and identical and the time direction was not distinguished. In contrast to that, in  CDT one assumes that the (Lorenzian) spacetimes are globally hyperbolic and thus they admit a time-foliation into spatial   hypersurfaces, called {\it  slices}, of equal cosmological time.
It is also required that the spatial topology $\Sigma$ of each slice  is fixed and preserved in the time evolution. As a result the spacetime  topology is that of  $[0,1]\times \Sigma$.   The (three-dimensional) geometry of a spatial slice at  time $t_k$ ($k$ denotes the lattice time coordinate, $k\in \mathbb{Z} $) is represented by a triangulation ${\cal T}^{(3)}_k$ (with the fixed topology $\Sigma$) constructed from identical equilateral 3-simplices (tetrahedra). Any two such triangulations  ${\cal T}^{(3)}_k$ at time $t_k$ and ${\cal T}^{(3)}_{k+1}$ at time $t_{k+1}$ constitute boundaries of a (four-dimensional) triangulation ${\cal T}^{(4)}_{k\, ;\, k+1}$ of a {\it slab} between $t_k$ and $t_{k+1}$, which can be constructed from four types of  4-simplices, denoted  $T^{(4,1)}$,  $T^{(1,4)}$, $T^{(3,2)}$ and $T^{(2,3)}$, where the numbers in parentheses represent the numbers of vertices  that each simplex has in time $t_k$ and $t_{k+1}$, respectively, see Fig. \ref{fig:simplices}. The 4-simplices are "glued" together in such a way that there are no topological defects in the slab and thus any spatial hypersurface in time $t\in(t_k , t_{k+1})$ will  also have the requested fixed topology $\Sigma$. The construction is  repeated for $k=1,2, ... k_{max}$ to obtain a full four-dimensional triangulation ${\cal T}$ with spatial boundaries ${\cal T}^{(3)}_1$ and ${\cal T}^{(3)}_{k_{max}}$. In practice, in order to avoid the necessity of defining spatial boundary conditions,  one usually identifies ${\cal T}^{(3)}_1$ and ${\cal T}^{(3)}_{k_{max}+1}$, i.e., one fixes  the spacetime topology  to be $S^1\times \Sigma$ instead of  $[0,1]\times \Sigma $.
The distinction of space and time introduced by the foliation is also present in the (fixed) edge lengths of the 4-simplices.  
This is quantified by  the  parameter $\alpha$, defined by
\beql{alpha}
   a_t^2 = -\alpha  \, a^2_s \, ,
\end{equation}
where $a_t$ and $a_s$ are the lattice spacing in the time and the spatial direction, respectively.
According to  Regge calculus \cite{regge}, the curvature of  a piece-wise flat simplicial manifold constructed from 4-simplices is defined through a deficit angle located at  two-dimensional subsimplices (triangles) and depends on  the number of 4-simplices sharing a given triangle. Using Regge prescription one can  compute the Einstein-Hilbert action for a CDT triangulation $\cal T$, the so-called Regge action  \cite{chapterJan,DynTriangLQG}
\beql{SR}
    S_{R}[k_0,k_4, \Delta; {\cal T}] = - \big(k_0 + 6\Delta\big) N_0 + k_4 \big(N_4^{(4,1)} + N_4^{(3,2)}\big) + \Delta N_4^{(4,1)}.
\end{equation}   
The action is a linear combination of $N_0$, denoting the total  number of vertices,  $N_4^{(4,1)}$,  the total number of  $T^{(4,1)}$ and $T^{(1,4)}$ 4-simplices, and $N_4^{(3,2)}$,  the total number of    $T^{(3,2)}$ and $T^{(2,3)}$ 4-simplices,  in the triangulation $\cal T$. The three dimensionless bare coupling constants are $k_0$, related to the  inverse  Newton's constant, $k_4$, related to the  cosmological constant, and $\Delta$, related to the  parameter $\alpha$ defined in equation \eqref{alpha}.

Using the CDT discretization the (formal) quantum-gravitational path integral \eqref{QGPI} can be defined by
\beql{CDTPI}
    {Z}^{(L)}_{CDT}(k_0,k_4, \Delta, \Sigma) =\sum_{\mathcal{T}\in \mathcal{T}^{(L)}} \frac{1}{{\cal C}_\mathcal{T}}e^{iS^{(L)}_R[k_0,k_4, \Delta; {\cal T}]},
\eeq
where the sum is over all possible CDT triangulations ${\cal T}$ 
described above, the measure term $1/{\cal C}_\mathcal{T}$ is the size of the automorphism group of $\mathcal{T}$ and  $S_R$ is the Regge action \eqref{SR}, and we have used the index $(L)$ to emphasize the Lorentzian setting. 
It is important to stress that despite the fact that in CDT one discretizes quantum geometries using triangulated manifolds with some finite lattice spacing $a$ it is not  assumed that the physical spacetime is discrete in any sense, and the UV cutoff $\sim 1/a$  (which regularizes the theory) is aimed to be removed in the continuum limit, if it exists. 
The  four-dimensional CDT, defined by the lattice regularized path integral \eqref{CDTPI}, cannot be solved analytically and in order to investigate its properties one is forced to use numerical  methods which require a change from the Lorentzian to the Euclidean setting. Due to the introduced time-foliation the metric signature change  is well defined. Wick rotation  can be achieved by an analytic continuation $\alpha\to -\alpha$ ($\alpha > 7/12$) in the lower half of the complex $\alpha$-plane. As a results the interior of each 4-simplex  becomes isomorphic to a (part  of) flat Euclidean space and the Regge action is changed accordingly such that $i \, S_R^{(L)} = - S_R^{(E)}$.\footnote{Both the Lorentzian $S_R^{(L)}$ and the Euclidean  $S_R^{(E)}$ Regge actions have the same form as that of  equation \eqref{SR}, just  the functional dependence of the CDT bare coupling constants $k_0,k_4, \Delta$ on $G, \Lambda, \alpha$ must be appropriately adjusted.} The Euclidean Regge action $S_R^{(E)}$ is purely real and thus the CDT path integral  \eqref{CDTPI} becomes a partition function 
\beql{CDTPF}
    {Z}^{(E)}_{CDT}(k_0,k_4, \Delta, \Sigma) =\sum_{\mathcal{T} \in \mathcal{T}^{(E)}} \frac{1}{{\cal C}_\mathcal{T}}e^{-S^{(E)}_R[k_0,k_4, \Delta; {\cal T}]}.
\eeq
Note that the  (Eucledean) triangulations $\cal T$ appearing in the above sum  must 
respect the imposed time-foliation  and the fixed spatial topology  $\Sigma$ of the original (Lorentzian) triangulations, thus a memory of the Lorentzian setting is still kept.

The partition function \eqref{CDTPF} can be investigated using  Monte Carlo (MC) methods. One 
performs MC simulations for a fixed topology $\Sigma$ and a choice of points in  the CDT bare coupling constants space $(k_0, k_4, \Delta)$. The results of such simulations show that for fixed values of $k_0$ and $\Delta$ the partition function behaves in the leading order as
\beql{K4crit}
    Z_{CDT}^{(E)}(k_0, k_4, \Delta, \Sigma)  \ \propto \ e^{(k_4^{c}-k_4) N_4} \ ,
\eeq 
where $N_4 = N_4^{(4,1)}+ N_4^{(3,2)} $. The factor $e^{-k_4 N_4}$ comes directly from the bare cosmological constant $k_4$ appearing in the Regge action (\ref{SR}).  The factor  $e^{k_4^{c} N_4}$ comes from the number of states (triangulations with $N_4$ 4-simplices) with the same value of the action, as  the number of possible configurations grows (approximately) exponentially with $N_4$, and the  exponent $k_4^{c}$ depends on $k_0$ and $\Delta$.
If $k_4<k_4^{c}$ the partition function is divergent and the CDT theory becomes ill-defined. For $k_4 > k_4^{c}$ the size of the system remains finite. Therefore taking the infinite volume limit ($N_4 \to \infty$) requires at the same time fine-tuning  $k_4\to k_4^{c}$. In the MC simulations it is convenient to let  the total lattice volume $N_4$ fluctuate around a fixed value $\bar N_4$ and perform a series of measurements for various target volumes $\bar N_4$.\footnote{In practice one usually fixes the $N_4^{(4,1)}$ volume.}  For each $\bar N_4$ the  $k_4$ coupling constant is fine-tuned  to the pseudo-critical value  $k_4\to k_4^{c}(\bar N_4)$. In the limit $\bar N_4\to \infty$ the finite-size effects vanish and one effectively obtains  $k_4 \to  k_4^{c}\equiv k_4^{c}(\infty)$. By fixing $\bar N_4$ one in fact studies the properties of $Z_{CDT}^{(E)}(k_0,\Delta, \bar N_4, \Sigma)$, which is linked with $Z_{CDT}^{(E)}(k_0, k_4, \Delta, \Sigma)$ by the Laplace transform
\beql{eq:ZCDTE}
    Z_{CDT}^{(E)}(k_0, k_4, \Delta, \Sigma)=\int_0^\infty d\bar N_4 e^{-k_4 \bar N_4}  Z_{CDT}^{(E)}(k_0,\Delta,\bar N_4, \Sigma)  .
\eeq
As a result, for the fixed topology $\Sigma$ and the target lattice volume $\bar N_4$, the partition function effectively depends  only on two bare coupling constants $k_0$ and  $\Delta$.
As will be shown below, at least in some region of the CDT parameter space, the dependence on $\bar N_4$ turns out to be universal and the $\bar N_4$ scaling analysis is a  useful tool in discussing the semiclassical and the (perspective) continuum limits of CDT.

\begin{figure}[ht!]
\centering
    {\includegraphics[width=0.45\textwidth]{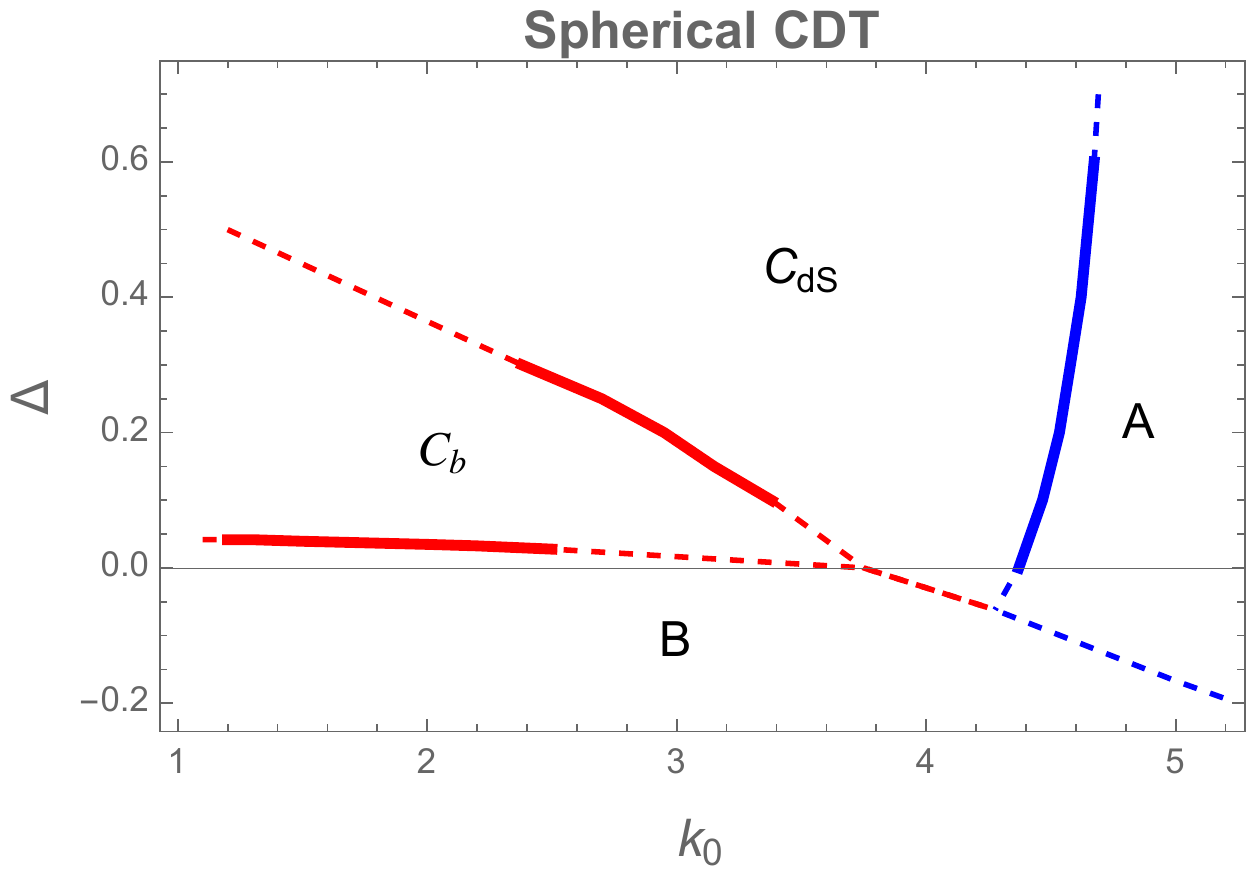}}
    {\includegraphics[width=0.45\textwidth]{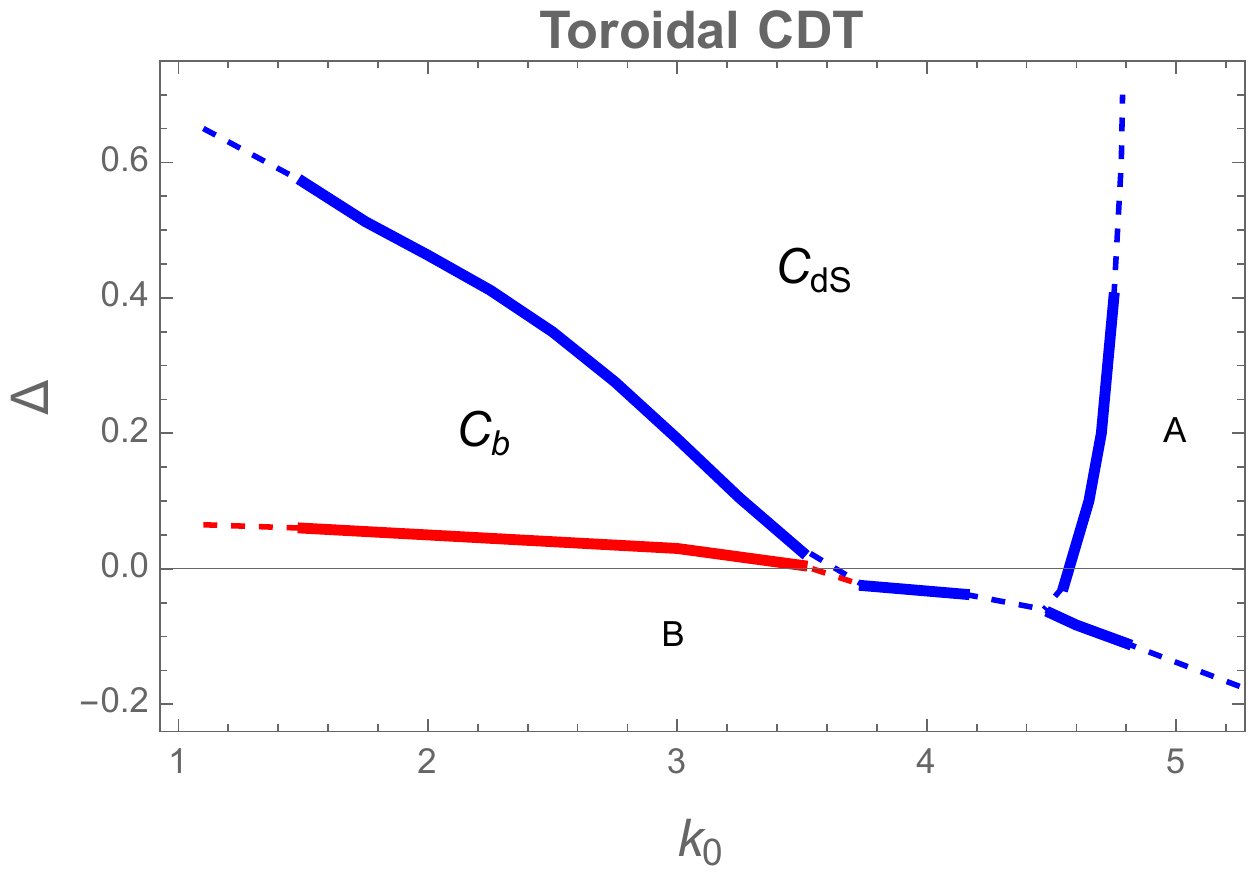}}
\caption{The phase diagram(s) of the spherical (left panel) and the toroidal (right panel) CDT. Solid lines denote measured phase transition lines, where first-order transitions are shown in blue while higher-order transitions in red. Dashed lines are extrapolations.}
\label{fig:phasediagram}
\end{figure}

\begin{figure}[ht!]
\centering
    \frame{\includegraphics[width=0.23\textwidth]{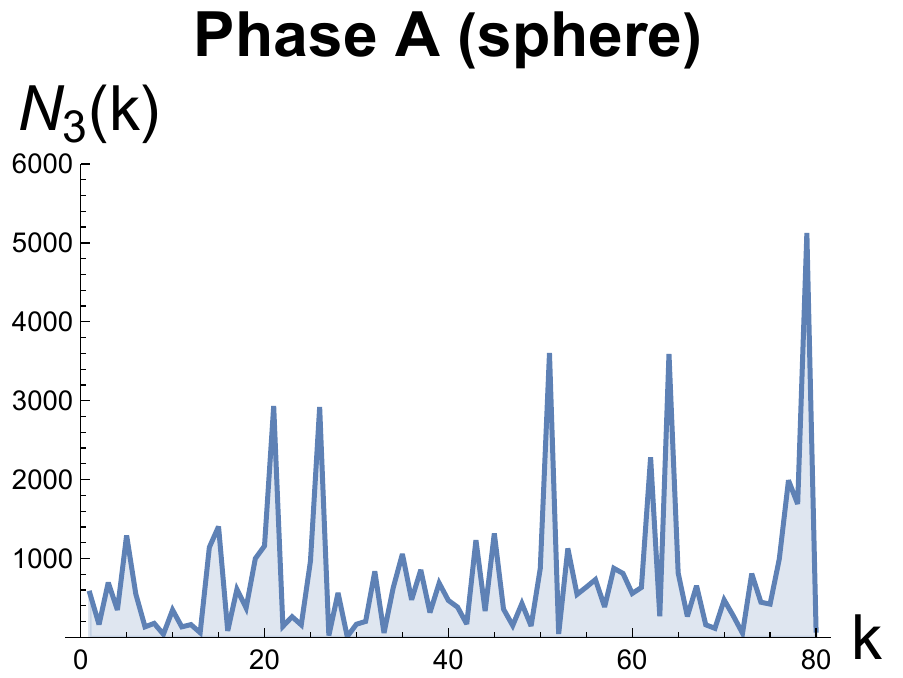}}
    \frame{\includegraphics[width=0.23\textwidth]{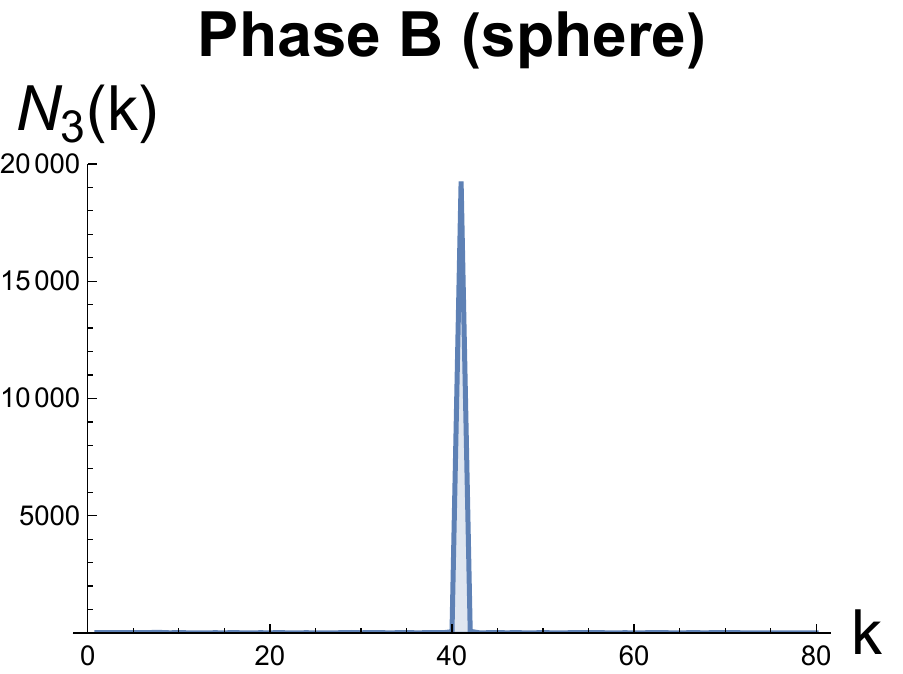}}
    \frame{\includegraphics[width=0.23\textwidth]{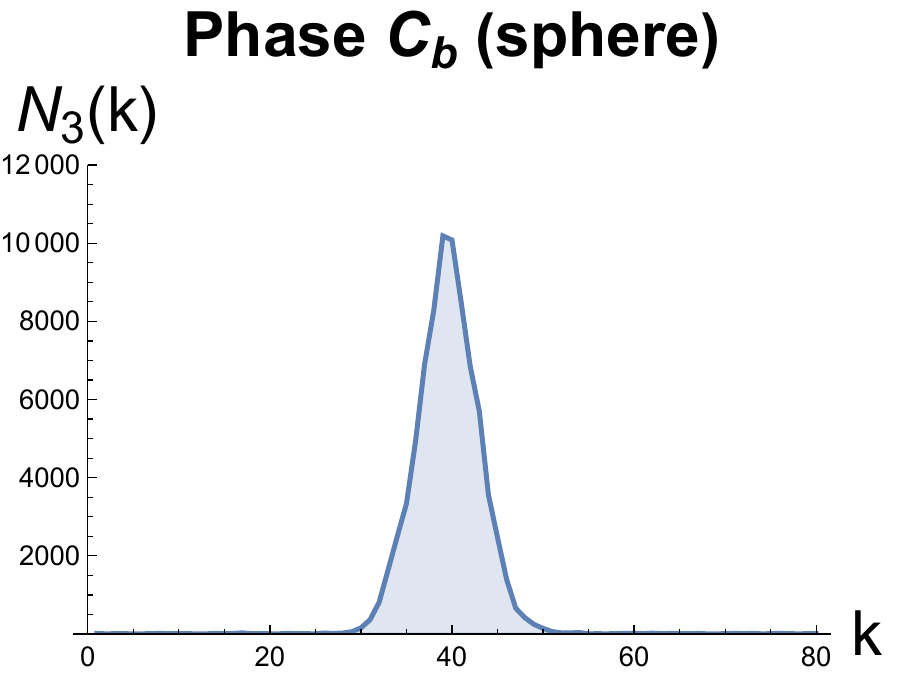}}
    \frame{\includegraphics[width=0.23\textwidth]{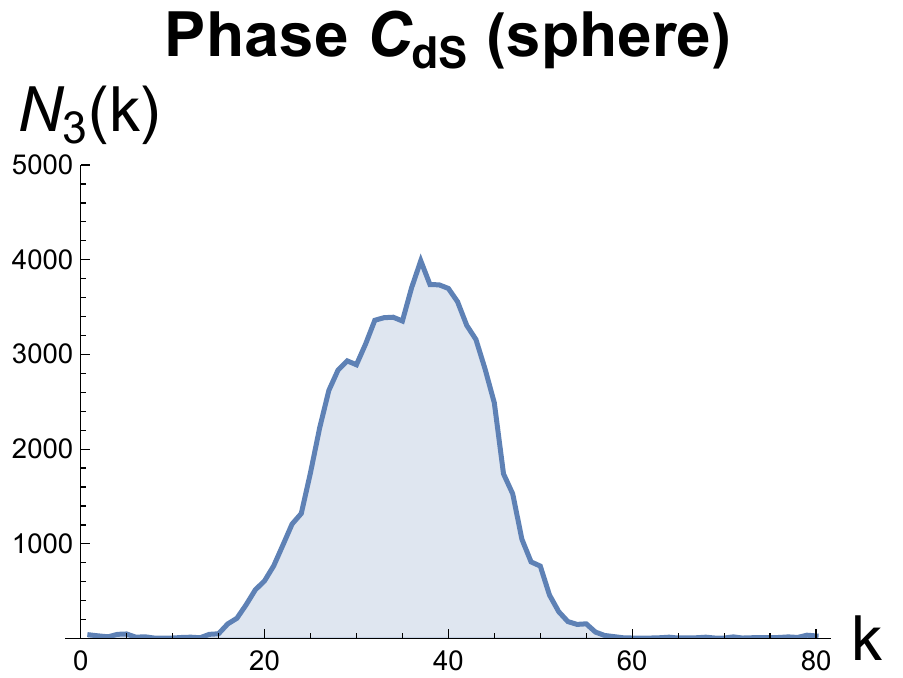}}\\
    \frame{\includegraphics[width=0.23\textwidth]{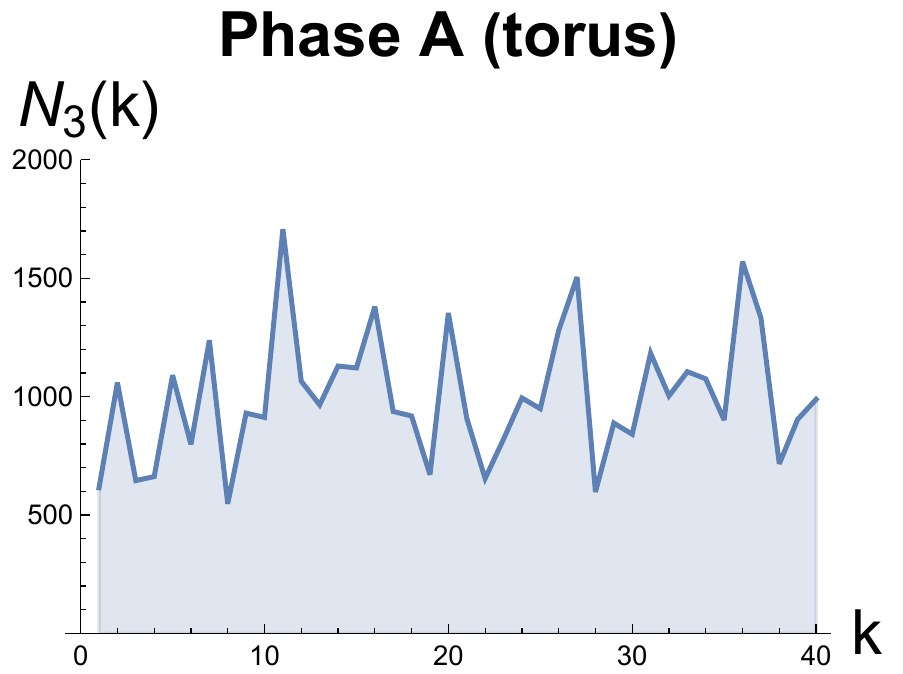}}
    \frame{\includegraphics[width=0.23\textwidth]{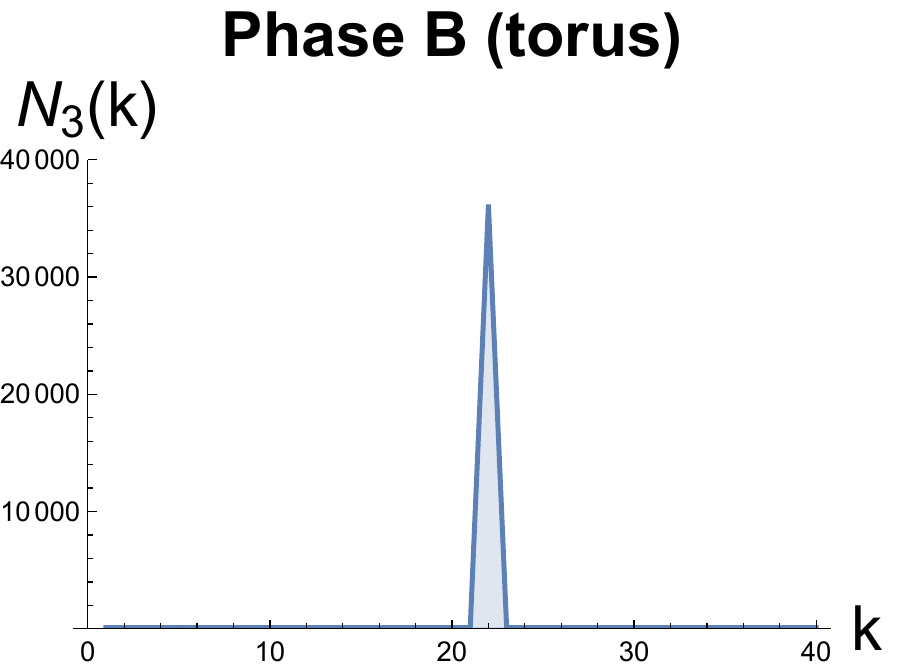}}
    \frame{\includegraphics[width=0.23\textwidth]{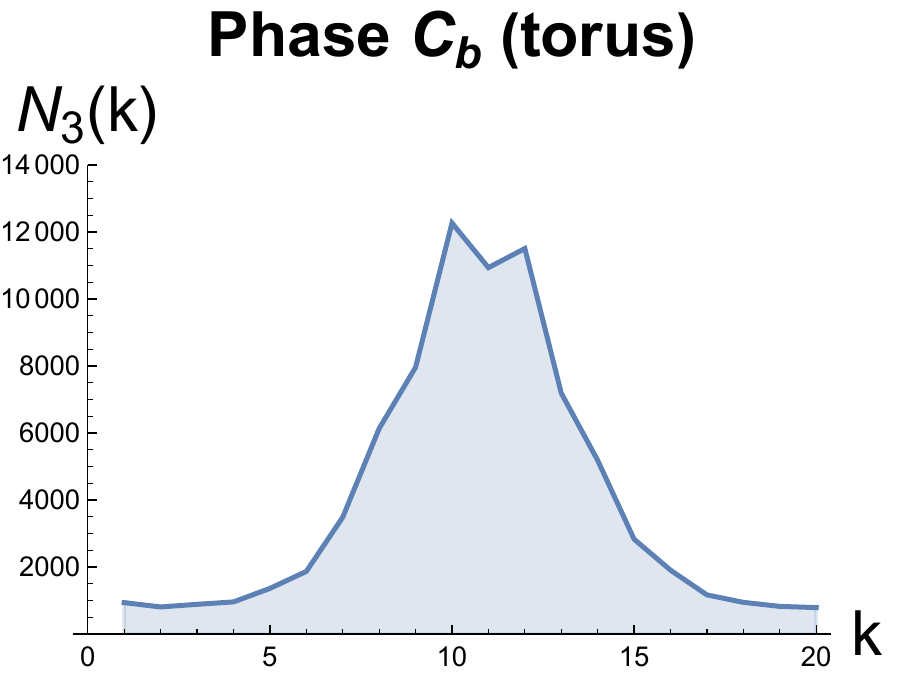}}
    \frame{\includegraphics[width=0.23\textwidth]{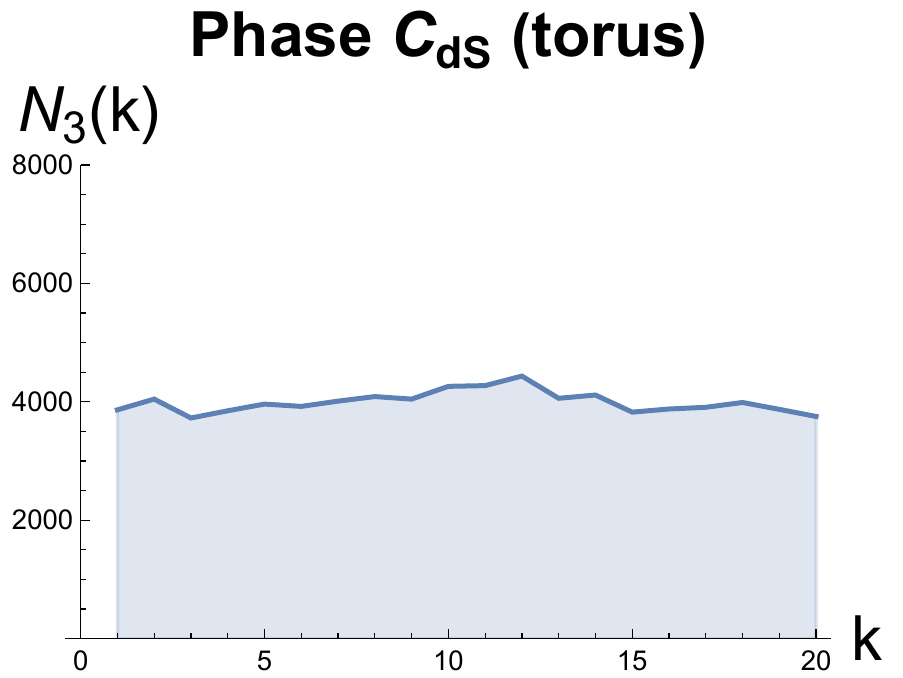}}\\
\caption{Spatial volume profiles $N_3(k)$ of generic CDT configurations in different phases. Top: spherical CDT: $A$, $B$, $C_b$, $C_{dS}$; Bottom: toroidal CDT:  $A$, $B$, $C_b$, $C_{dS}$, respectively.}
\label{fig:phases}
\end{figure}

The parameter space of the four-dimensional CDT spanned by $k_0$ and $\Delta$  has been largely investigated for the cases where the spatial topology $\Sigma$ was chosen to be either $S^3$ (the, so-called, {\it spherical CDT}) or $T^3$ (the {\it toroidal CDT}). Four phases with distinct geometric properties have been found \cite{CDTHL, Signature, PhasestructureTorus}, see Fig.  \ref{fig:phasediagram}.   
Phase $A$ is observed for sufficiently large values of the  bare (inverse) cosmological constant $k_0$.  A typical configuration consists of  many disjoint "baby universes"  with a short time extension, see Fig. \ref{fig:phases}. In phase $A$ the three-dimensional spatial volumes, quantified by the number of  tetrahedra $N_3(k)$ forming a spatial slice with the lattice time coordinate $k\in \mathbb{Z}$, are uncorrelated. 
Phase $B$ is realized  for small values of the bare coupling $\Delta$. 
In contrast to  phase $A$, inside  phase $B$ the whole geometry "collapses" into a single spatial slice containing almost all spatial volume. The slice ends in the "past" and the "future"  in a vertex of very high coordination number (belonging to almost all 4-simplices). The spatial volume outside the collapsed slice is close to the cutoff size, see Fig. \ref{fig:phases}. 
It is believed that these two phases can be understood essentially from the dynamics of the three-dimensional EDT geometries of the slices of constant (lattice) time. The spatial slices in phase $A$ are presumably realizations of branched polymers which dominate the path integral of three-dimensional Euclidean quantum gravity for large $k_0$, whereas the extended part of the geometry in phase $B$ corresponds to the Euclidean geometry in the, so-called, crumpled phase, which is characterized by the presence of very few vertices with extremely high coordination number.
The  important feature of CDT, earlier unobserved in EDT is the existence  of two new phases of quantum geometry.
For sufficiently small values of $\kappa_0$ and positive, but not too-large $\Delta$ one can observe   phase $C_{b}$. It shares some features with phase $B$, namely it contains vertices of very large coordination number, but, at the same time, it has a time-extended geometry, see Fig. \ref{fig:phases}. The presence of high-order vertices surrounded by volume clusters means that the geometry is quite inhomogeneous. Finally, for even higher values of the bare coupling $\Delta$ one reaches phase $C_{dS}$, whose  generic geometry depends on the spatial topology choice, see Fig. \ref{fig:phases}. In the spherical CDT the geometry consists of the time-extended {\it blob}, where most three-volume is placed, and the spatial volume changes  quite smoothly with the time coordinate, whereas in the toroidal CDT one observes a constant volume profile. The observed shape of the quantum universe closely resembles classical cosmological  solutions of General Relativity. Therefore the phase $C_{dS}$ can be considered to be related to the semiclasscial limit of CDT.

\section{The semiclasscial limit}

As already mentioned in the introduction, 
in the four-dimensional CDT one  obtains new phases of quantum geometry, earlier unobserved in EDT. 
The especially interesting one is the phase $C_{dS}$, also called the {\it de Sitter} or the {\it semiclassical} phase. The hallmark of this phase is the presence of a stable extended four-dimensional geometry,
where on large scales, compared to the triangulations size, a  homogeneous and isotropic semiclassical spacetime with superimposed quantum fluctuations emerges dynamically. As will be discussed below, in phase $C_{dS}$ not only does the effective macroscopic dimension of spacetime emerges with the correct classical value, but, equally remarkably, the global shape of the universe has been shown to be related to a simple minisuperspace description, similar to that used in standard quantum cosmology.

\subsection{Evidence for the effective spacetime dimension four}

It is important to stress that, in the path-integral formulation, 
the notion of spacetime dimension is an emergent concept. This is related to the fact that  geometries entering the gravitational path-integral are 
not that of  smooth four-dimensional manifolds
and therefore the effective dimension can, and in general does, differ from the topological dimension.  In the lattice regularized models, such as CDT, it exemplified by the fact that even though the 4-simplices forming triangulations are four-dimensional objects, due to the non-trivial connectivity structure, the triangulations cannot be  embedded in the four-dimensional space. One should therefore define some notion of an effective dimension, which can be different (both larger or smaller) from the topological dimension.

\subsubsection{The Hausdorff dimension}

\begin{figure}[ht!]
\centering
    \includegraphics[width = 0.45\textwidth]{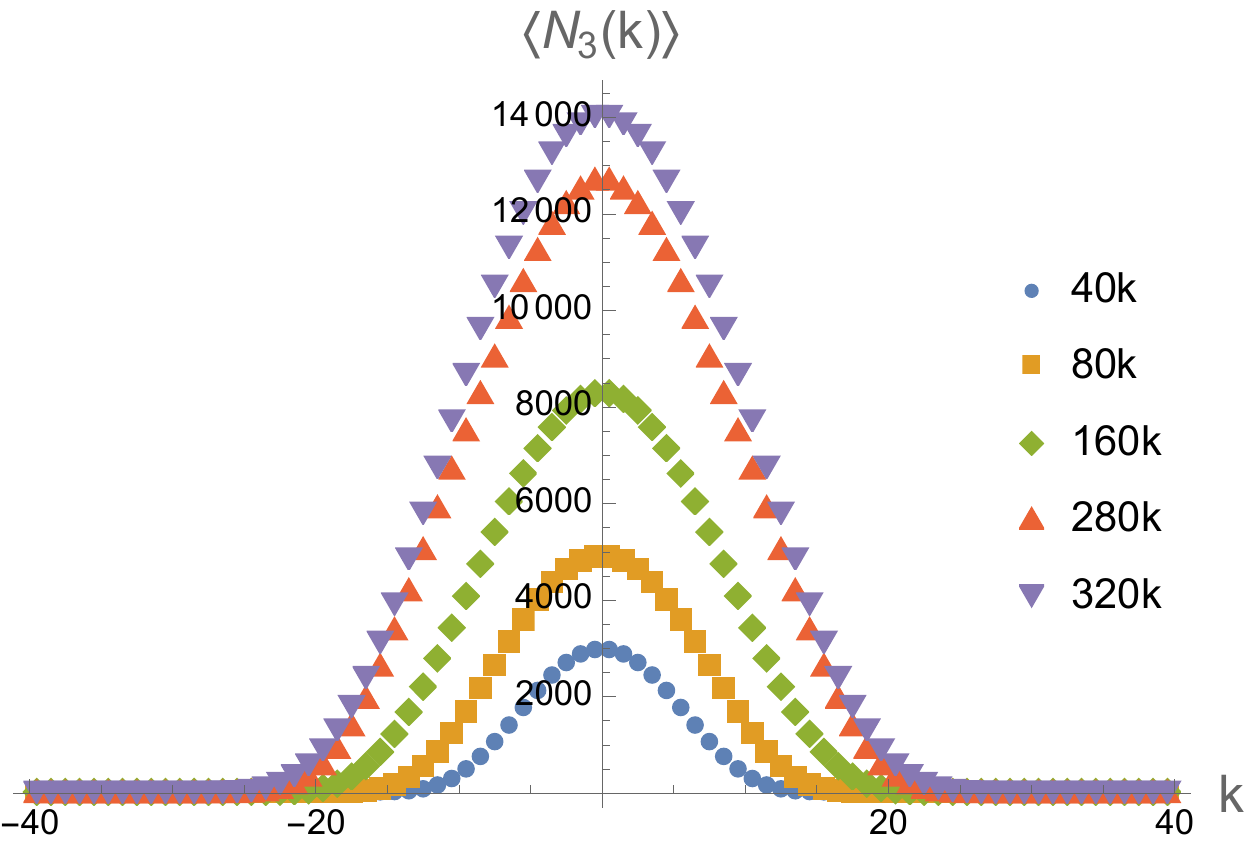}
    \includegraphics[width = 0.45\textwidth]{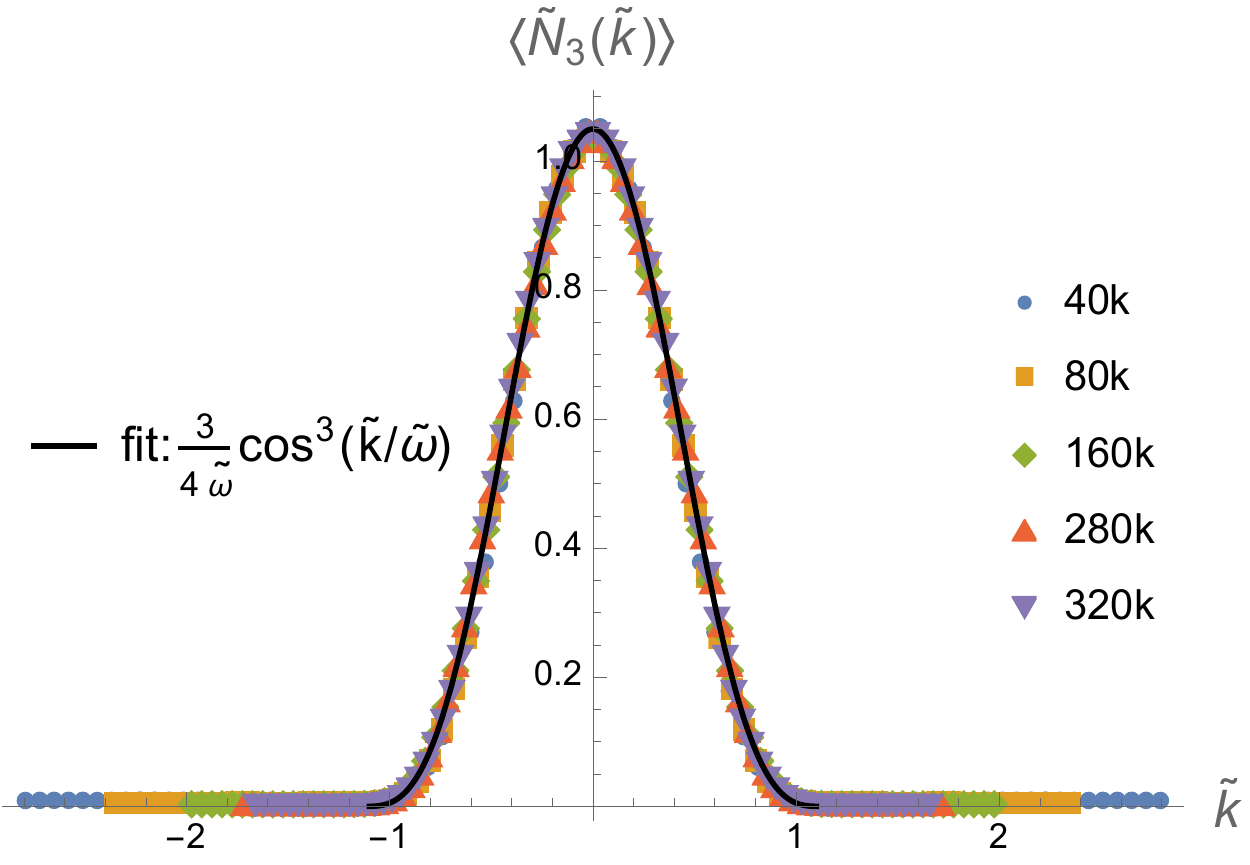}
\caption{Left panel: average volume profiles $\langle N_3(k) \rangle_{\bar N_4}$ measured inside phase $C_{dS}$ of the spherical CDT ($k_0=2.2, \Delta=0.6)$. Different colors denote different target lattice volumes $\bar N_4^{(4,1)}$. Right panel: the volume profiles rescaled according to equation \eqref{eq:rescale} with Hausdorff dimension $d_H=4$ and the fit of equation \eqref{eq:deSitter11}.}
\label{fig:vol_prof}
\end{figure}

Due to the imposed time-foliation, one of the simplest observables\footnote{Here, by an {\it observable} we mean a quantity measured in the numerical simulations, not a  diffeomorphism invariant observable.} measured in  the Monte Carlo simulations of CDT is the so-called {\it spatial volume profile} $N_3(k)$. Each spatial slice with the lattice time coordinate $k\in \mathbb{Z}$ and  the corresponding (continuous) time coordinate $t_k = k \, a$, where $a$ is the lattice spacing\footnote{In the following we will assume for simplicity that the effective lattice spacing $a$ is the same in time and spatial directions. One can generalize to different lattice spacing $a_t$ and $a_s$.}, is a three-dimensional triangulation constructed from identical equilateral tetrahedra. 
Accordingly,  the continuum three-volume $V_3(t_k)$  
is  proportional to the number $N_3(k)$ of spatial tetrahedra in $k$
\beql{eq:V3volume}
    V_3(t_k) \propto a^3 N_3(k). 
\eeq
In the {\it spherical} CDT, the generic spatial volume profile observed in phase $C_{dS}$  consists of an extended part, the so-called {\it blob}, whose ends are connected by a thin {\it stalk} where the spatial volume is of the minimal allowed cutoff size, see Fig. \ref{fig:phases}. 
During a MC simulation, due to the CDT time-translation symmetry, the blob performs a slow random walk around the (periodic) time-axis. Therefore, if one performed a long-enough MC simulation\footnote{In practice such a simulation would take several years of computer time.} the  spatial volume profile averaged over configurations would be  a constant, in accordance with the time-translation symmetry. At the same time it is obvious that the presence of the blob means that individual CDT configurations in phase $C_{dS}$ break the time-translation symmetry. 
One can take this into account when computing the average over configurations by appropriately  
shifting the center of  volume   of each individual MC generated configuration to a common time coordinate, say  $k=0$. 
As a result one gets a non-trivial  one-point correlator, i.e., the {\it average volume profile} $\langle N_3(k) \rangle_{\bar N_4}$, see Fig. \ref{fig:vol_prof}, where the index $\bar N_4$ denotes the constant target lattice  volume fixed in a MC simulation.

 If the effective spacetime dimension is $d_H$ then, using textbook scaling arguments,  one should expect that the spatially extended parts of the spacetimes measured for various effective four-volumes $\tilde N_4= \langle \sum_{k \in \text{blob}} N_3(k) \rangle_{\bar N_4}$, where the sum is over the extended part of the universe, can be mapped onto each other by rescaling the proper-time and the spatial volumes  according to
\beql{eq:rescale}
    k \to \tilde k =\frac{k}{\tilde N_{4}^{1/d_H}}  \quad , \quad N_3(k) \to \tilde N_3(\tilde k)=\frac{N_3(k)}{\tilde N_{4}^{1-1/d_H}}.
\eeq
As a result  
\beql{eq:resvolprof}
    \langle N_3(k) \rangle_{\bar N_4}  = \tilde N_4^{1-1/d_H} \langle \tilde N_3(\tilde k) \rangle = \tilde N_4^{1-1/d_H}\, H \left(\frac{k}{\tilde N_{4}^{1/d_H}}\right )
\eeq
can be expressed in terms of a universal function $H$. 
The  {\it Hausdorff dimension} $d_H$ can be then measured by performing a series of MC measurements  for systems of various size $\tilde N_4$ and fitting equation \eqref{eq:resvolprof} by applying the best overlap method. In Fig.~\ref{fig:vol_prof} we show that indeed, to a  good approximation, our MC generated data measured  in phase $C_{dS}$ of the   spherical CDT fit  the above scaling relation  with  a classical value of Hausdorff dimension $d_H=4$.

\subsubsection{The spectral dimension}

Another way of obtaining an effective dimension of  spacetime appearing in a non-perturbative quantum gravity is to study a diffusion process of a fictitious point particle, related to the, 
  so-called, {\it spectral dimension}, for more details see also \cite{chapterWlosi}.\footnote{The diffusion process discussed here is related to the  {\it scalar} spectral dimension, one can also consider a more general diffusion of $k$-forms, see  \cite{Marcus}.}  Let us consider a  smooth $d$-dimensional Riemannian manifold ${\cal M}$ equipped with a metric $g_{ij}$. The diffusion process is governed by the  equation
\beql{eq:diffusion}
\frac{\partial}{\partial \sigma}K(x,x_0;\sigma) = \Delta_g K(x,x_0;\sigma),
\eeq
where $\sigma$ is the fictitious diffusion time and $\Delta_g$ is the Laplace-Beltrami operator of the metric $g_{ij}$,
with the initial condition peaked at $x_0 \in  \mathcal{M}$, i.e., given by a delta function $K(x, x_0; \sigma = 0) = \delta^d(x - x_0)$.
The  heat kernel $K(x,x_0;\sigma)$ represents the probability density  of finding a particle at position $x\in \mathcal{M}$ after  a diffusion time $\sigma$.  The average return probability to the origin after  time $\sigma$
\beql{eq:returnprob}
    P_r(\sigma) = \frac{1}{V}\int dx \sqrt{g} K(x,x;\sigma),
\eeq
where $V = \int dx \sqrt{g}$ is the volume of $\mathcal{M}$, has the following asymptotic expansion for small $\sigma$ \cite{Craioveanu}
\beql{eq:retprobexpansion}
    P_r(\tau) \sim \sigma ^{-\frac{d}{2}}\sum_{i=0}^{\infty} A_i\sigma^i,
\eeq
where $A_i$ are generally complicated integrated functions of the metric $g_{ij}$.
Note that the return probability \eqref{eq:retprobexpansion} depends on the dimension $d$ of the manifold. One can therefore compute the dimension from the return probability 
\beql{eq:dimspec}
    d=-2 \lim_{\sigma \to 0} \frac{d \log P_r(\sigma)}{d \log \sigma}.
\eeq
The expression \eqref{eq:dimspec} can be generalized by dropping the limit $\sigma \to 0$ which defines the (running) {\it spectral dimension}
\beql{eq:specdim}
    d_s(\sigma)=-2  \frac{d \log P_r(\sigma)}{d \log \sigma}.
\eeq

\begin{figure}[ht]
\centering
    \includegraphics[width = 0.6\textwidth]{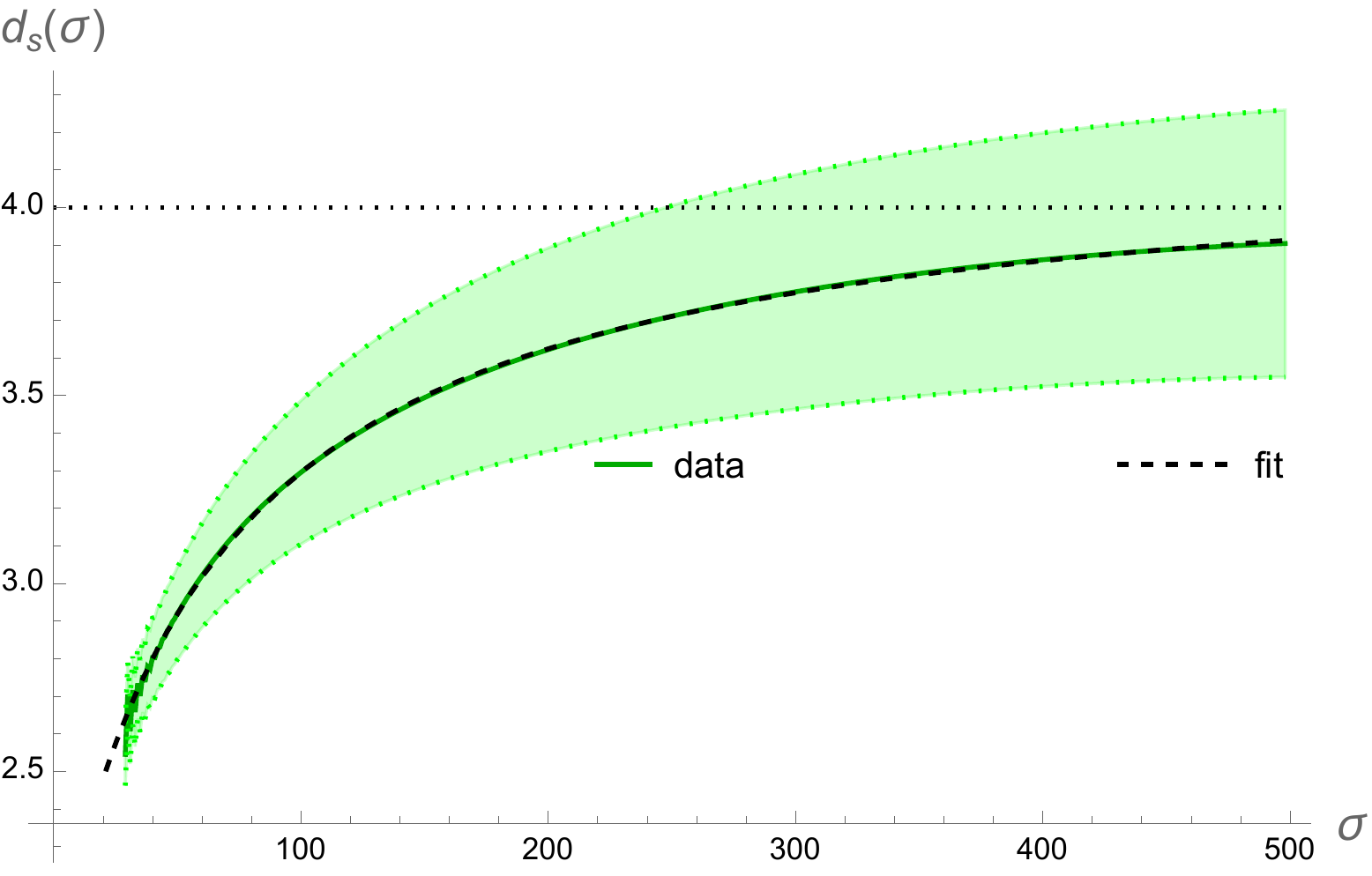}
\caption{The spectral dimension $d_s(\sigma)$ measured inside  phase $C_{dS}$ of the spherical CDT ($k_0=2.2, \Delta=0.6)$ and the fit of a phenomenological function \eqref{eq:retprobfit}. Shaded region indicates measurement errors.}
\label{fig:spectral_dim}
\end{figure}

In CDT one deals with discretized geometries and therefore one also has to discretize the diffusion equation \eqref{eq:diffusion}. One typically considers the discrete diffusion on the {\it dual lattice}, i.e., the graph whose nodes are 4-simplices and whose links encode neighborhood relations between the 4-simplices.   The diffusion process is originated at a randomly chosen 4-simplex of a given triangulation $\mathcal{T}$ and the return probability  is computed for a range of discrete diffusion steps $\sigma=0, ..., \sigma_{max}$. Then the (manifold) average return probability $ P_r(\sigma) $ is estimated  over a number of starting  points $x_0\in \mathcal{T}$. As we are interested in the expectation value of the above observable,   the average  $\langle P_r(\sigma) \rangle_{\bar N_4} $ over the ensemble of triangulations is subsequently computed.  Finally,  the (expectation value of) spectral dimensions is calculated using equation \eqref{eq:specdim}. The results obtained in phase $C_{dS}$ of the spherical CDT 
are presented in Fig.~\ref{fig:spectral_dim}, where we also included a phenomenological fit 
\beql{eq:retprobfit}
    d_s(\sigma)=d + \frac{c}{\sigma+b},
\eeq
where the constants $d=4.17\pm0.07,c=146\pm28,b=66\pm17$ were established from data in range $\sigma\in[30,450]$.
As can be seen the  spectral dimension changes  with the diffusion time $\sigma$, or otherwise with the geodesic distance,  as the diffusion process probes larger and larger scales of the quantum geometry.\footnote{One should  note that for very large diffusion times $\sigma$, due to  finite size effects, i.e., when the diffusion process probes the whole simplicial manifold (which is by construction closed in the chosen spacetime topology $S^1 \times S^3$), the zero modes of the Laplace-Beltrami operator dominate and the spectral dimension $d_s(\sigma)$  goes to zero.  This part of the evolution is not shown on the plot and was not taken into account when fitting equation \eqref{eq:retprobfit}.} 
The remarkable feature of the curve $d_s(\sigma)$ is its slow approach to the asymptotic value of $d_s(\sigma=\infty)=4.17\pm0.07$ consistent  with  the  four-dimensional universe. This type of behaviour is non-trivial and is not the case in other CDT phases.  The other non-trivial phenomenon that has emerged dynamically 
is a scale dependence of the spectral dimension  which goes to  $d_s(\sigma=0)\approx 2$ for short distances. It is related to  non-classical features of generic geometries appearing in the CDT path integral. This non-trivial "dimensional reduction" has been first found in \cite{SpectralDimCDT} and confirmed in  many other approaches to quantum gravity \cite{AS49, NCG47, HLG48, LQG52}.

Summarizing this part, both the Hausdorff and the spectral dimensions measured in the  $C_{dS}$ phase  point to the dynamically emerging macroscopic four-dimensional "average geometry". In the following  we will provide evidence that indeed the geometry is consistent with a semiclassical four-dimensional universe.

\subsection{Evidence for an effective minisuperspace action}\label{sec:2}

\subsubsection{The semiclassical de Sitter universe}

In the lattice formulation one is  interested in taking the thermodynamical (infinite volume) limit $\bar N_4\to \infty$, where one may hope to recover  continuum physics. As already mentioned, inside phase $C_{dS}$ of  the spherical CDT the one-point correlator, i.e., the average spatial volume, shows a non-trivial scaling behaviour consistent with  Hausdorff dimension four. One  can therefore define a universal function $H(\tilde k)$, describing the (rescaled) volume profile, 
see equation \eqref{eq:resvolprof}, which is independent of the lattice volume and thus it is also valid  in the thermodynamical limit. A very nontrivial fact is that it 
is very well fitted by (cf. Fig. \ref{fig:vol_prof})
\beql{eq:deSitter11}
    H(\tilde k) =\frac{3}{4 \,\tilde \omega}\cos^3\left( \frac{\tilde k}{\tilde \omega} \right),
\eeq
 resulting in the universal scaling
\beql{eq:deSitter}
    \langle N_3(k) \rangle_{\bar N_4}  =  \frac{3  \tilde N_4 }{4}\frac{1}{ \tilde \omega  \tilde N_4^{1/4} } \cos^3\left( \frac{ k}{\tilde  \omega  \tilde N_4^{1/4}} \right).
\eeq
The above is of course valid only  inside the extended part of the universe (the blob), i.e., for $-\frac{\pi}{2} \tilde \omega  \tilde N_4^{1/4} \leq  k \leq \frac{\pi}{2} \tilde \omega  \tilde N_4^{1/4} $. 
The constant $\tilde  \omega$ appearing in equation \eqref{eq:deSitter} is a function on the CDT bare couplings $k_0$ and $\Delta$.  
Restoring continuum  quantities one can write 
\begin{eqnarray}\label{eq:deSitterV3}
    \langle V_3(t) \rangle  &=& \frac{3 V_4}{4}  \frac{1}{ \omega  V_4^{1/4}}  \cos^3\left(  \frac{ t}{ \omega V_4^{1/4} } \right) = \nonumber \\
    &=& 2 \pi^2 {\cal R}^3\cos^3\left(\frac{ t} {{\cal R}} \right),
\end{eqnarray}
which is   the volume profile of a regular four-sphere  with  physical volume $V_4$ and radius ${\cal R }=  \omega V_4^{1/4}$, and  where $ t$ is the (Euclidean) {\it proper-time}. It  corresponds to the Wick rotated (Euclidean) de Sitter universe, i.e., a vacuum  solution of Einstein's field equations with positive cosmological constant $\Lambda = 3 / {\cal R }^2 $. Such a solution is obtained for 
 a maximally symmetric (spatially homogeneous and isotropic) metric with the infinitesimal line element squared 
\beql{eq:metric}
    ds^2= dt^2+a^2(t)d\Omega_3^2 \ ,
\eeq 
where $a(t)$ is the scale factor and $d\Omega_3^2$ denotes the line
element squared on the unit three-sphere, such that the spatial volume at proper-time $t$ is $V_3(t) = 2 \pi^2 a^3(t)$.

\subsubsection{The effective action from volume fluctuations}

So far we have looked only at the one-point function, i.e., the expected value of the spatial volume observable 
\beql{eq:volscale}
    \langle N_3(k)\rangle_{\bar N_4} = \tilde N_4^{3/4}\,   H \left(\frac{k}{ \tilde \omega \tilde N_{4}^{1/4}}\right) .
\eeq

One can as well measure the two-point (connected) correlator
\beql{eq:correl}
    \langle \delta N_3(k)  \delta N_3(k') \rangle_{\bar N_4},
\eeq
where
\beql{eq:fluct}
    \delta N_3(k) = N_3(k) - \langle N_3(k) \rangle_{\bar N_4}
\eeq
denote fluctuations around the dynamically emerging semiclassical average. 

Let us first comment that the correlator \eqref{eq:correl} again shows the following universal scaling
\beql{eq:Corrscale}
    \langle \delta N_3(k)  \delta N_3(k') \rangle_{\bar N_4} = \tilde N_4 \, F\left(\frac{k}{\tilde  \omega \tilde N_{4}^{1/4}}, \frac{k'}{\tilde  \omega \tilde N_{4}^{1/4}}\right)
\eeq
with the same $\tilde \omega$, as in equation \eqref{eq:volscale}.
The details of  the function $F$ 
are not important for the following discussion. Comparing equations \eqref{eq:volscale} and \eqref{eq:correl} implies that the (relative amplitudes of) quantum fluctuations scale as
\beql{eq:dV}
    \frac{\sqrt{\langle \delta V_3^{\, 2}(t_k) \rangle}}{\langle V_3(t_k) \rangle}=\frac{\sqrt{\langle \delta N^{\, 2}_3(k) \rangle_{\bar N_4}}}{\langle N_3(k)\rangle_{\bar N_4} }\propto \frac{1}{\tilde N_4^{1/4}},
\eeq
i.e., they vanish in the thermodynamical limit $\bar N_4 \to \infty$. As a result one obtains a fully classical spatial volume trajectory consistent with the de Sitter solution \eqref{eq:deSitterV3}. Nevertheless, as will be shown below, analysis of  quantum fluctuations described by the two-point correlator \eqref{eq:Corrscale} is a useful tool to determine  the  {\it effective action} of CDT.

Let us consider a continuous model of quantum fluctuations  around a well defined semiclassical trajectory $\bar n(t)\equiv \langle n(t) \rangle$. In a semiclassical  approximation 
the quantum fluctuations $\delta n(t)$ around this trajectory are described by a Hermitian operator $P(t,t')$ obtained by a quadratic expansion of the effective action
\beql{SExpCont}
    S_{eff}\big[\bar n(t) + \delta n(t)\big] = S_{eff}\big[\bar n(t)\big] + \frac{1}{2}\int dt\,dt'\ \delta n(t) P(t,t')  \delta n(t')   ,
\eeq
where we neglect all terms of order ${O}(\delta n ^3)$ in fluctuations.
In a discretized model
\beql{Sexpansion}
    S_{eff}\big[\bar n(t_k) + \delta n(t_k)\big]  =  S_{eff}\big[\bar n(t_k)\big]+ \frac{1}{2} \sum_{k k'} \delta n(t_k) P_{k k'} \delta n(t_{k'})
\eeq
$P$ becomes a matrix parametrized by a (discrete) time variable $k$
\beql{Ptt}
    P_{k k'}= \left.\frac{\partial^2 S_{eff}}{\partial n(t_k)\partial n(t_{k'})}\right|_{n(t)=\bar n(t)}.
\eeq
 For  such an expansion quantum fluctuations  
 are Gaussian and the (discrete) two-point correlator, i.e., the covariance matrix, is given by the inverse of the propagator 
\begin{equation}
    C_{kk'}\equiv \langle \delta n(t_k)  \delta n(t_{k'}) \rangle  =  \left(P^{-1}\right)_{k k'} \ .
\eeq 
 The problem can simply be  inverted. In numerical simulations  one can measure the covariance matrix $ C$. The inverse of the $C$ matrix defines  the (propagator) $ P$ matrix  and thus the effective action $S_{eff}$, or at least its  derivatives measured at the semiclassical solution (see equation \eqref{Ptt}). 
 
 In order to apply the above method in the case of CDT, one should first check whether the quantum fluctuations are sufficiently well described by a quadratic effective action implying Gaussian fluctuations. This is indeed the case for the observed  spatial volume fluctuations measured in the extended part of the CDT universe (the blob) in phase $C_{dS}$ of the spherical CDT. 
 Therefore the method seems applicable in the physically interesting limit of large lattice volumes. The  empirical  matrix $\hat P$ , obtained by inverting the measured spatial volume-volume correlator $\hat C$
 has a simple tridiagonal structure, 
 suggesting that there are only 
 nearest neighbour interactions between  adjacent  spatial slices. 
 Consequently, the effective action can be expressed in the following form
\beql{SeffGeneral}
    S_{eff} = S_{kin}+S_{pot} =\sum_k  K\Big(N_3(k),N_3(k+1)\Big) + \sum_k  V\Big(N_3(k)\Big) \ ,
\eeq
where the functions $K$ and $V$, representing respectively  the kinetic  and the potential terms, have to be determined from empirical  data. 
In order to do so let us make an Ansatz about a possible  form of the effective action.
As already mentioned, the semiclassical spatial volume profile 
is consistent with the de Sitter spacetime with spherical spatial topology, obtained for  the 
metric \eqref{eq:metric}. 
Inserting the metric  
to the (Euclidean) Einstein-Hilbert action
\beql{CSHA}
    S_{HE}^{(E)}=-\frac{1}{16 \pi G}\int{d t}\int{d\Omega_3 \sqrt{g}(R-2 \Lambda)} \ 
\eeq
 one obtains the  following {\it minisuperspace} action
\beql{CS2}
    S_{MS}^{(S^3)}=-\frac{3 \pi}{4  G} \int{d t }  \left(  a\dot a^2 + a -\frac{\Lambda}{3} a^3\right).
\eeq
The Euler-Lagrange equation of motion for the scale factor $a(t)$ derived from the  action \eqref{CS2} is solved by
\beql{Csola}
    \bar a(t) = {\cal R} \cos\left(\frac{ t} {{\cal R}} \right) ,
\eeq
where we have chosen  the proper-time  to be in  range  $-\frac{\pi}{2} {\cal R}\leq  t\leq \frac{\pi}{2} {\cal R} $, such that the maximum is located at $t=0$.
Reparametrizing $a(t)$ in terms of  the spatial volume
\beql{CSV}
    V_3(t) = \int d\Omega_3\sqrt{g|_{S^3}}=2\pi^2 a^3(t)
\eeq
one indeed obtains a solution consistent with the average CDT solution \eqref{eq:deSitterV3}
\begin{eqnarray}\label{CsolMSV}
    \bar V_3(t) &=& 2 \pi^2 {\cal R}^3\cos^3\left(\frac{t} {{\cal R}} \right)= \nonumber \\
    &=& \frac{3 V_4}{4}  \frac{1}{ \omega  V_4^{1/4}}  \cos^3\left( \frac{t}{ \omega V_4^{1/4} } \right)  \quad , \quad  V_4 = \int_{-\frac{\pi}{2} {\cal R}}^{\frac{\pi}{2} {\cal R}} dt \, \bar V_3(t).
\end{eqnarray} 
The constant $ \omega = (\frac{3}{8 \pi^2})^{1/4}$ is set by a condition that   the total four-volume of the sphere is equal $V_4$. 
The simplest discretization of the above continuum expression is  given by 
\beql{eq:deSitter2}
    \bar  N_3(k)  =  \frac{3  \tilde N_4 }{4}\frac{1}{ \tilde \omega  \tilde N_4^{1/4} } \cos^3\left( \frac{ k}{\tilde  \omega  \tilde N_4^{1/4}} \right) \quad , \quad \tilde N_4 = \sum_k \bar N_3(k).
\eeq
Comparing equations \eqref{CsolMSV} and \eqref{eq:deSitter2} one can identify 
\beql{identification1}
  \frac{t_k}{\omega V_4^{1/4}}  =  \frac{k}{\tilde\omega \tilde N_4^{1/4}} \quad , \quad V_4 = C_4 \tilde N_4 a^4,
\eeq
where the (dimensionless) constant $C_4$ is the volume of a unit 4-simplex, 
 $a$ is the lattice spacing, and the continuum proper-time $t_k = k \, a$, implying
\beql{identification2}
   \tilde \omega = \omega  \, C_4^{1/4} .
\eeq

The minisuperspace action \eqref{CS2} can  also be expressed in terms of the spatial volume \eqref{CSV} 
\beql{CSMSV}
    S_{MS}^{(S^3)}=-\frac{1}{ \Gamma}\int dt  \left( \frac{ \dot V_3^2}{V_3}+  \mu V_3^{1/3}- \lambda V_3\right)  ,
\eeq
where $ \Gamma = 24 \pi G$, $ \mu =  9 \left( 2 \pi^2\right)^{2/3}$ and $ \lambda = 3 \Lambda$.
The simplest  discretization of the above action has, up to a sign\footnote{Note that the effective action of CDT agrees with the minisuperspace action up to an overall sign, see Conclusions for more discussion of this fact.}, the following form
\beql{eq:Sdiscrete}
    S_{eff}^{(S^3)}
    =\frac{1}{ \tilde \Gamma} \sum_k  \Bigg( {\Bigg. 2 \frac{\left( N_3(k)-N_3(k+1)\right)^2}{  N_3(k)+N_3(k+1)}} +  {\Bigg.\tilde  \mu \, N_3(k)^{1/3} -\tilde  \lambda \, N_3(k) }  \Bigg)  ,
\eeq
which is the Ansatz we were looking for.
 The  empirical propagator matrix 
$\hat P$, measured inside phase $C_{dS}$ of the spherical CDT, 
is indeed very well described by the proposed Ansatz  \cite{NonperturbativeQdSU,SemiclassicalJGS}.  
The dimensionless  coupling constants $\tilde  \Gamma, \tilde  \mu$ and $\tilde  \lambda$ appearing in the discretized effective action \eqref{eq:Sdiscrete} can be fitted from numerical data.
Similarly to $\tilde \omega$ 
they all depend on the 
CDT bare couplings $k_0$ and $\Delta$. 

By analyzing  relations between dimensionful and dimensionless quantities appearing respectively in the continuum expression   
\eqref{CSMSV} and its discrete counterpart \eqref{eq:Sdiscrete} one obtains 
\begin{eqnarray}
    \label{identification3}
    \sqrt{C_4} \, a^2\, \tilde \Gamma \tilde \omega^2  & = &  \Gamma  \omega^2, \\ 
    \label{identification3a}
    \tilde \mu\,\tilde  \omega^{8/3} & = &   \mu\,   \omega^{8/3},  \\
     \label{identification3b}
    {\tilde \lambda \, \tilde \omega^2} & = &  {\sqrt{C_4} \, a^2} \, \lambda \, \omega^2.
\end{eqnarray}
Rewriting  \eqref{identification3} in terms of the Newton  constant  ($\Gamma=24 \pi G\, ;\,  \omega = (\frac{3}{8 \pi^2})^{1/4}$) leads to 
\beql{eq:GGamma}
    a^2={\frac{6 \sqrt{6} }{\sqrt{C_4} \tilde \Gamma \tilde  \omega^2}} G.
\eeq

Assuming that the (renormalized) Newton Constant  $G$ is fixed and, as suggested by the CDT results, the semiclassical description is still valid close to the Planck length  $\ell_{p}=\sqrt{G} $ (in units where $c=\hbar=1$), one can use equation \eqref{eq:GGamma} to estimate the effective lattice spacing (in $[\ell_p]$).
The lattice spacing depends on two scales: the average size of the CDT universe expressed in lattice units (quantified by $ \tilde  \omega$) and the amplitude of quantum fluctuations (quantified by $ \sqrt{\tilde \Gamma}$). The lattice spacing at the "generic" point in phase $C_{dS}$ ($k_0=2.2, \Delta=0.6$), where most of the CDT simulations were done, is estimated to be $a \approx 2 \; \ell_{p}$ which translates into an average radius of the MC generated quantum geometries  $\sim 10 \; \ell_{p}$. It is remarkable that quantum universe at such short scales can be so well described by a semiclassical minisuperspace approximation.

\subsubsection{The effective action from transfer matrix method}

As argued above, the spatial  volume distribution (or alternatively the scale factor) of  the spherical CDT inside the $S_{dS}$ phase is very well described by a simple discretization of the minisuperspace action \eqref{eq:Sdiscrete}. This result was obtained  by analyzing a semiclassical approximation of volume fluctuations around the average (de Sitter) volume profile, where  one could use the empirical covariance matrix to compute the propagator matrix. The problem with such an approach is that in a semiclassical approximation the propagator matrix is defined by second derivatives of the effective action at the semiclassical solution. One should note that the form of the discretized effective action \eqref{eq:Sdiscrete} suggests that, in the physically interesting blob region, the second derivatives, i.e., the elements of the propagator matrix, fall very fast with increasing volume (derivatives of the kinetic term behave as $N_3(k)^{-1}$ and of the potential term as $N_3(k)^{-5/3}$). As a result it is very difficult to observe any subleading corrections, especially in the effective potential. 
A natural question arises if one can measure the effective action without resorting to the semiclassical approximation thus validating the above results.

The quasi-local form of a CDT discretized effective action \eqref{SeffGeneral} leads in a natural way to the path-integral representation of the spatial volume fluctuations
\beql{Zeff}
    {Z}_{eff} =\sum_{{\{N_3(k)\}}}e^{-S_{eff}}\equiv \sum_{{\{N_3(k)\}}} \prod_{k=1,...,k_{max}} e^{-L_{eff}[N_3(k),N_3(k+1)]} =  \text{tr}  { M}^{k_{max}} \ ,
\eeq
where  ${ M}$ is the {\it effective transfer matrix} defined by the effective Lagrangian 
\beql{TMLag}
    \langle{N_3(k+1)} | { M } | {N_3(k)} \rangle \propto  \exp\big (-L_{eff}[N_3(k),N_3(k+1)] \big).
\eeq
In the partition function \eqref{Zeff} one assumes time-periodic boundary conditions, consistent with  the CDT numerical setup.
In such approach one disregards all details of the geometric structure at a given spatial slice  and looks only at the spatial volume  $N_3(k)$. 

One should note that, by construction, 
CDT  has a {\it genuine  transfer matrix} ${\cal M}$ which relates a spatial geometry ${\cal T}^{(3)}_k$ at the lattice time $k$ to a  spatial geometry ${\cal T}^{(3)}_{k+1}$ at time $k+1$. 
The transfer matrix ${\cal M}$ is   defined by  a sum over all possible four-dimensional triangulations ${\cal T}^{(4)}_{k\, ;\, k+1}$ of a slab between $k$ and $k+1$ (with boundary three-dimensional triangulations ${\cal T}^{(3)}_k$ and ${\cal T}^{(3)}_{k+1}$)
\beql{ja2}
    \big \langle { {\cal T}^{(3)}_{k+1}| {\cal M} | {\cal T}^{(3)}_{k}} \big \rangle =
    \sum_{{\cal T}^{(4)}_{k\, ;\, k+1}} \ e^{-S_R[{\cal T}^{(4)}_{k\, ;\, k+1}]} \ ,
\eeq
where ${S_R[{\cal T}^{(4)}_{k\, ;\, k+1}]}$ is the Regge action computed for the four-dimensional triangulation of the slab. 
Accordingly, the partition function of CDT \eqref{CDTPF} corresponding to $k_{max}$ time steps (with time-periodic boundary conditions) can  be expressed as
\beql{ZFullCDT}
    {Z_{CDT}} = \sum_{\{{\cal T}_k^{(3)}\} }  \prod_{k=1,...,k_{max}} \langle{\cal T}^{(3)}_{k+1}  | {\cal M} |{\cal T}_{k}^{(3)}\rangle=\text{tr} {\cal M}^{k_{max}} .
\eeq
The effective transfer matrix elements can  be formally computed as an average over  the genuine transfer matrix elements 
\beql{Mnm}
    \langle  N_3(k+1)|{M}|N_3(k)\rangle= \frac{1}{{{\cal N}_{ N_3(k+1)} {\cal N}_{N_3(k)}}} 
    \sum_{\substack{{\cal T}^{(3)}_{k+1} \in {\cal T}^{(3)}(N_3(k+1))\\ {\cal T}^{(3)}_{k} \in{\cal T}^{(3)}(N_3(k))}} 
    \; \langle{\cal T}^{(3)}_{k+1} | {\cal M}|{\cal T}^{(3)}_k\rangle \ ,    
\eeq
where ${\cal T}^{(3)}(N_3(k))$ denotes a subset of all three-dimensional triangulations with size $N_3(k)$ and the number of such triangulations equals ${\cal N}_{N_3(k)}$. 
The statement that one can use the matrix $M$ defined above as an 
effective transfer matrix assumes that the standard deviation
of the matrix elements $\langle{\cal T}^{(3)}_{k+1} | {\cal M}|{\cal T}^{(3)}_{k}\rangle $ in equation \eqref{Mnm}  is sufficiently small and consequently
\beql{consistency}
    \langle{\cal T}^{(3)}_{k+\Delta k}| {\cal M}^{\Delta k} |{\cal T}^{(3)}_k\rangle\sim \langle  N_3(k+\Delta k)|{ M}^{\Delta k}|N_3(k)\rangle,
\eeq
for $\Delta k = 1,2,...k_{max}$.
In the following we will assume that the description of spatial volume fluctuations using the effective transfer matrix ${M}$  is a sufficiently good approximation and use it to analyze the empirical (Monte Carlo generated) data. The consistency of such analysis will provide indirect evidence that the approach is correct. 

Using the  partition function \eqref{Zeff} and the resulting effective transfer matrix $M$ one can compute the  
(two-point) probability distribution of measuring a combination of  volumes $N_3(k)=n\, , \, N_3(k')=m$ ($n,m\in \mathbb{N}$) in spatial slices separated by  $\Delta k = k' - k$ time steps
\beql{P2nt}
    P^{k_{max}}_{\Delta k}(N_3(k)=n\, ,\, N_3(k')=m) = \frac{1}{{ Z}_{eff}} \ {\langle{n | {M}^{k_{max}-\Delta k} | m\rangle} \langle{m| {M}^{\Delta k} | n}\rangle }\ ,
\eeq
where we again assumed time-periodic boundary conditions.
Let us consider a CDT model with a very short time period $k_{max}=2$.  It follows from equation \eqref{P2nt} that  the two-point probability distribution  of measuring $N_3(1)=n$ and $N_3(2)=m$  is given by
\beql{P2T2}
    P^{k_{max}=2}_{\Delta k=1}(N_3(1)=n,N_3(2)=m) = \frac{ \ {\langle{n | { M} |m}\rangle \langle{m | { M} |n}\rangle }}{{\text{tr} {M}^2}}\ .
\eeq
Due to   time reflection symmetry the transfer matrix $ M$ must be symmetric, leading to
\beql{MT2}
    \langle{n|{ M}|m}\rangle ={\cal N}_0 \sqrt{ P^{k_{max}=2}_{\Delta k=1}(N_3({1})=n,N_3({2})=m)} .
\eeq
The empirical two-point probability distribution $\hat P^{k_{max}=2}_{\Delta k=1}$ can be measured in  the MC simulations of CDT and therefore used to estimate the transfer matrix elements (up to a normalization factor ${\cal N}_0$), see Fig.~\ref{fig:transfer}, where
 we show (the logarithm) of the  transfer matrix elements measured 
 in phase $C_{dS}$ of the spherical CDT.  
The empirical transfer matrix can be then used to reconstruct the effective Lagrangian $L_{eff}$. According to equation \eqref{TMLag} one gets
\beql{eq:Leff}
    L_{eff}[n,m]=-\log\langle{n|{\cal M}_{eff}|m}\rangle + \log {\cal N}_0 
\eeq
and the corresponding effective action is given by
\beql{eq:SeffTM}
    S_{eff}= \sum_k L_{eff}[N_3(k), N_3(k+1)].
\eeq
For very small spatial volumes ($N_3<\sim 300$) the effective Lagrangian is dominated by lattice discretization artifacts (Fig.~\ref{fig:transfer}, left), but for larger volumes it is quite smooth (Fig.~\ref{fig:transfer}, right) and can be very well fitted by the following (discretized) Lagrangian \cite{TMJGS}
\beql{LeffCC}
    L_{eff}^{(S^3)}[n,m] = \frac{1}{\tilde  \Gamma}\left[ 2 \frac{(n-m)^2}{n+m-2 n_0} + \tilde \mu \left(\frac{n+m}{2}\right)^{1/3} - \tilde  \lambda \left(\frac{n+m}{2}\right) \right] \, ,
\eeq
where we have omitted a constant term $\log {\cal N}_0$.
Fitted values of the parameters $\tilde \Gamma, \tilde \mu$ and $\tilde \lambda$ agree within measurement errors with the parameters obtained using the covariance matrix method discussed above. In the above discretization we have also included a (very small\footnote{Fitted values of $|n_0| < 10$ are very small compared to macroscopic spatial volumes $N_3 > 300$, where discretization effects are negligible and the parametrization \eqref{LeffCC} holds.}) finite-size correction $n_0$. Note that in \eqref{LeffCC}  we have also used a different discretization of the  potential part of the effective Lagrangian than in equation \eqref{eq:Sdiscrete}. Such a form better fits the transfer matrix data but it gives exactly the same continuum effective action, consistent with the minisuperspace  action \eqref{CSMSV}.

\begin{figure}[ht]
\centering
    \includegraphics[width = 0.45\textwidth]{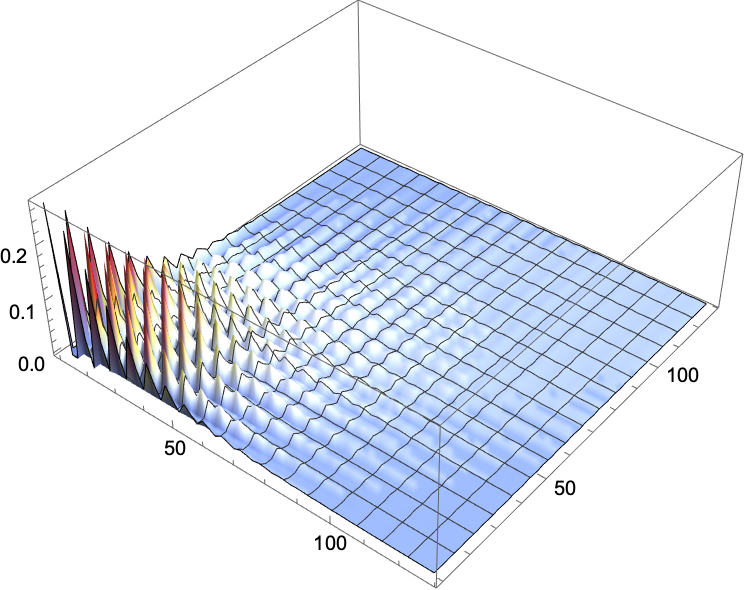}
     \includegraphics[width = 0.45\textwidth]{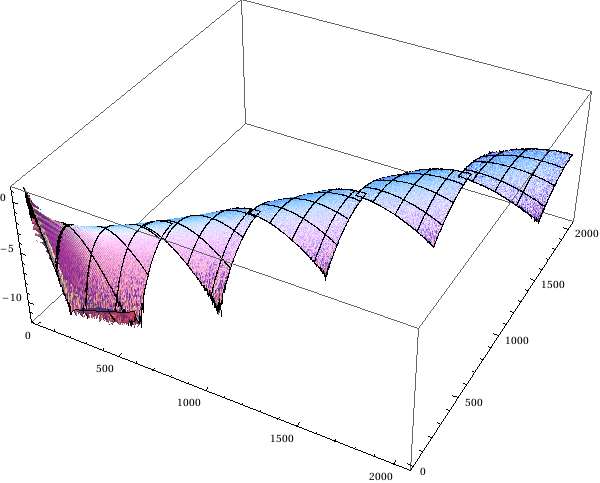}
\caption{Empirical transfer matrix measured for small (left panel) and large (right panel) spatial volumes in phase $C_{dS}$ of the spherical CDT ($k_0=2.2, \Delta=0.6)$. Right figure is in logarithmic scale.}
\label{fig:transfer}
\end{figure}

In order to check the validity of the effective transfer matrix approach one can compare the spatial volume fluctuations data measured in CDT and in a simplified effective model, defined by the partition function \eqref{Zeff}, where the spatial volume is the only degree of freedom. The transfer matrix ${M}$ used in the effective model is obtained by smoothly joining the purely empirical transfer matrix ${\hat  M}$, measured for very small spatial volumes where  dicretization artifacts prevail,  with the theoretical transfer matrix $M_{th}^{(S^3)}\equiv {\cal N}_0\exp  \left(- L_{eff}^{(S^3)} \right)$ defined by the effective Lagrangian \eqref{LeffCC} (with fitted values of $\tilde \Gamma, \tilde \mu$ and $\tilde \lambda$), applicable to larger volumes, i.e.,
\beql{Mextr}
    \langle n| { M}|m \rangle = \left\{ \begin{array}{ll}
    \ \ \langle n|{ \hat M}|m \rangle &, \ n < thr \textrm{ or } m < thr\\
    \langle n|  M_{th}^{(S^3)}|m \rangle &, \ \textrm{otherwise}
    \end{array} \right.,
\eeq
where $thr$ is a threshold ($thr = 300$).
In Fig. \ref{fig:effectivemodel} we compare the average spatial volume profile and the fluctuations profile measured in both the effective and the original CDT Monte Carlo simulations \cite{EffactionJGS}. The agreement is indeed remarkable thus validating the form of the effective action.

 \begin{figure}[ht]
\centering
    \includegraphics[width = 0.45\textwidth]{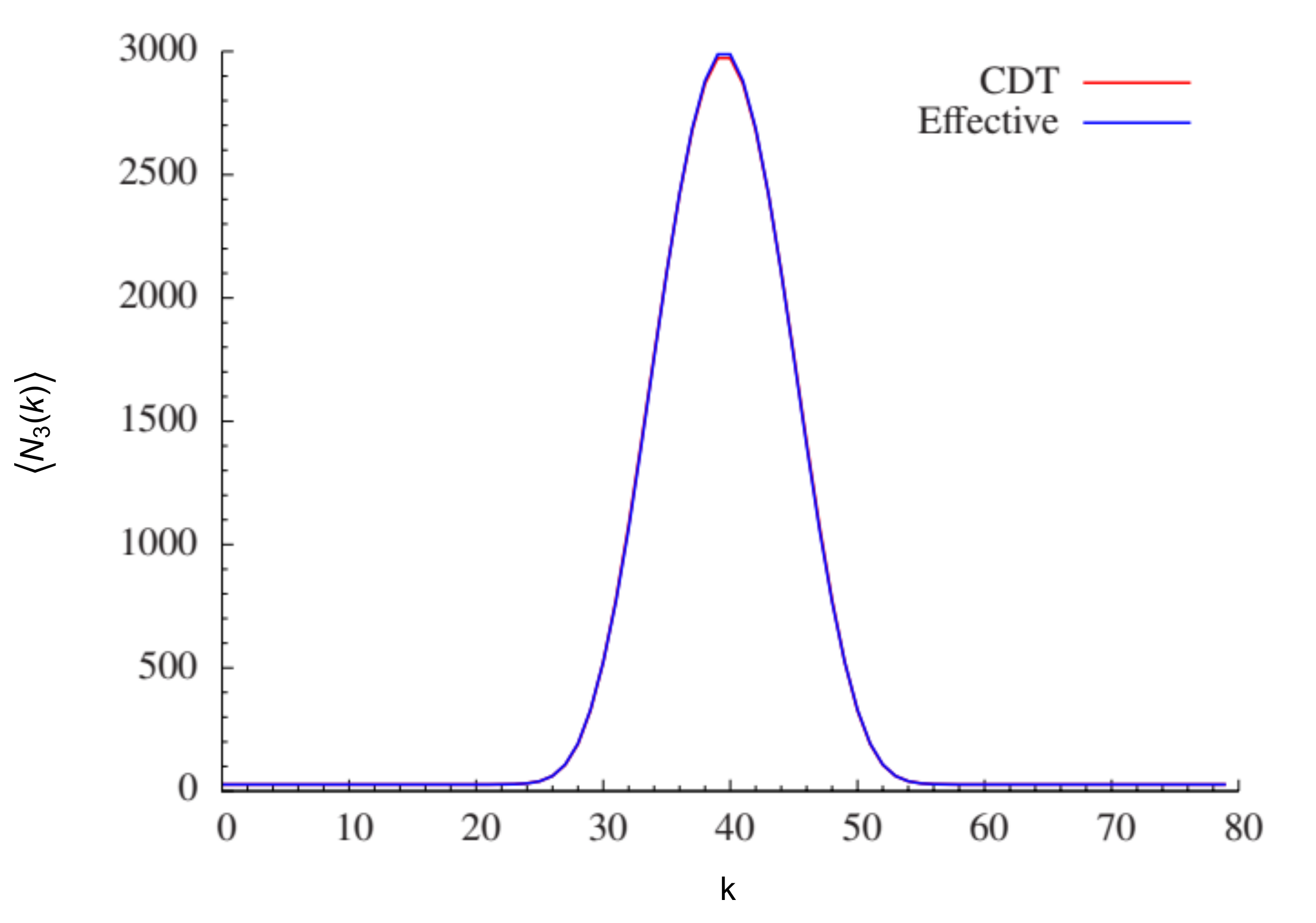}
     \includegraphics[width = 0.45\textwidth]{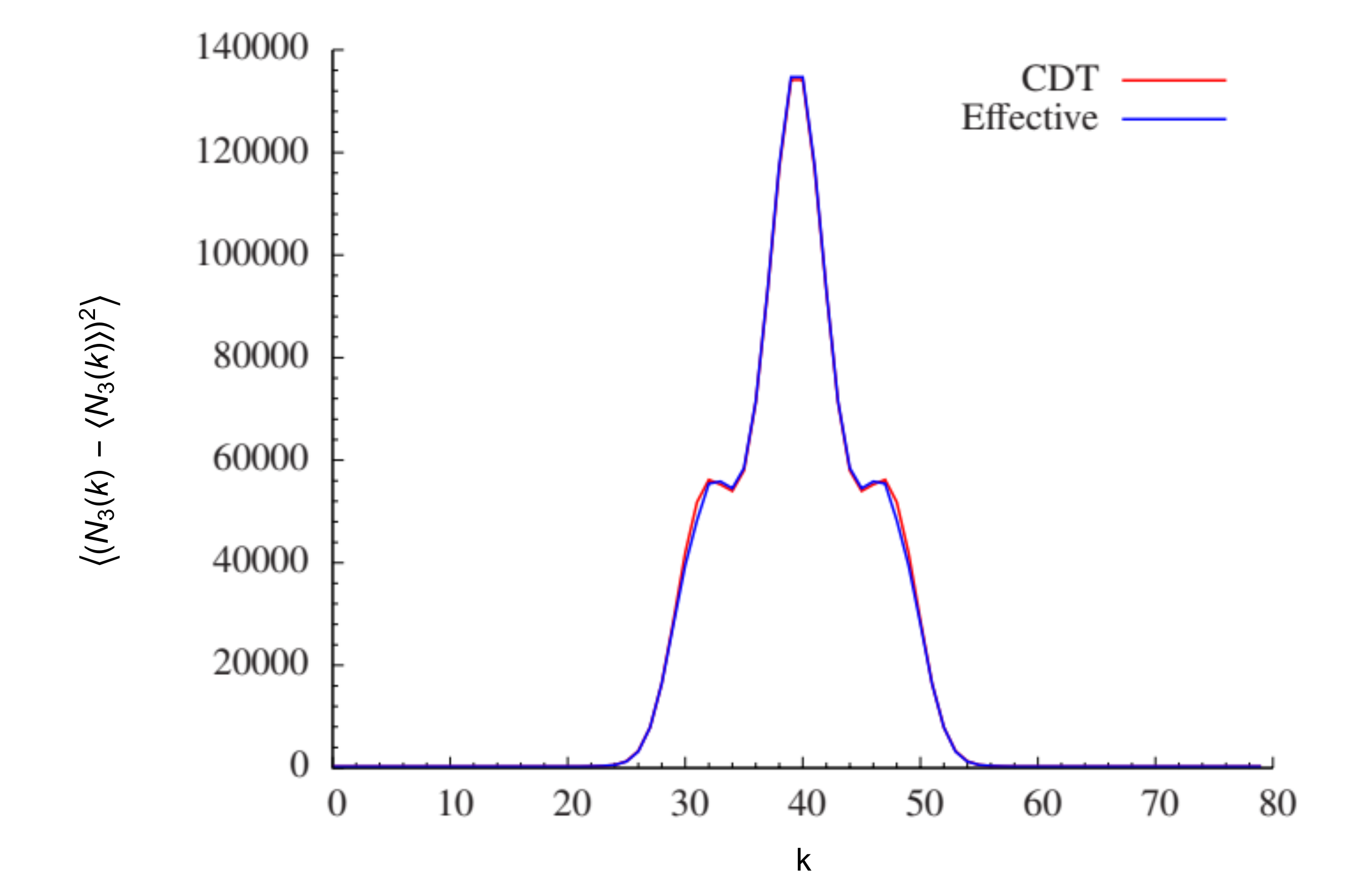}
\caption{Comparison of the  full CDT results with the effective model of spatial volume fluctuations described in the text. Left panel shows the average volume profile $\langle N_3(k) \rangle$ and right panel the fluctuations profile $\big\langle \big(N_3(k) - \langle N_3(k) \rangle \big)^2 \big\rangle $. Data  measured in phase $C_{dS}$ of the spherical CDT ($k_0=2.2, \Delta=0.6$).}
\label{fig:effectivemodel}
\end{figure}

\subsubsection{Impact of the spatial topology choice}

All the results discussed above were obtained in the CDT model with a (fixed) spherical spatial topology $\Sigma=S^3$. As discussed in the introduction, the spatial topology choice is an important ingredient of CDT, it is therefore important to check the impact of the topology choice on the  results.    
So far, the only other spatial topology  studied in the MC simulations was the toroidal topology $\Sigma=T^3\equiv S^1\times S^1 \times S^1$, and such a topology change (from spherical to toroidal) does not seem have impact on the CDT phase structure, see Fig. \ref{fig:phasediagram}.  All four phases ($A$, $B$, $C_b$ and $C_{dS}$) earlier observed in the spherical CDT are also present in the toroidal CDT. Also location of  all phase transition lines is similar for both topologies.\footnote{As will be discussed later, the order of the phase transitions can be different.} 
Nevertheless, due to the different spatial topologies the dynamically emerging  background geometry observed in phase $C_{dS}$ can be different. 
It is indeed the case as can be visualized by the spatial volume profiles measured in phase $C_{dS}$, see Fig. \ref{fig:phases}. In the spherical CDT the average volume profile was very well fitted by a $\cos^3$ function, cf. equation \eqref{eq:deSitter}, while in the toroidal CDT one rather observes  a constant volume profile  $\langle N_3(k) \rangle_{\bar N_4} = \text{const.}$ 

As discussed above, in the spherical CDT the spatial volume fluctuations are very accurately explained by the minisuperspace effective action, see equation \eqref{CSMSV}, where the potential term 
$\propto V_3(t)^{1/3}$ comes from a (constant positive) curvature of the three-sphere and therefore is absent in the case of the (flat) topological torus.
One can therefore expect that the toroidal version of phase $C_{dS}$ should admit the 
following minisuperspace action
\beql{CSMSVT}
    S^{(T^3)}_{MS}=-\frac{1}{24 \pi G}\int dt  \left( \frac{ \dot V_3^2}{V_3}- \lambda V_3\right) \, ,
\eeq
where the above potential term is absent. 

Using the covariance matrix of spatial volume fluctuations 
or the effective transfer matrix approach one can again determine the form and parameters of the  (discretized) effective action in the toroidal CDT, which is very well described by \cite{torus1,torus2}
\beql{LeffCTorus}
    S_{eff}^{(T^3)}
    =\frac{1}{\tilde  \Gamma} \sum_k  \Bigg( 2 {\frac{\left( N_3(k)-N_3(k+1)\right)^2}{  N_3(k)+N_3(k+1)}} -  {  \tilde  \lambda \, N_3(k) + \ \tilde  \xi \, N_3(k)^{\rho}}  \Bigg) \, ,
\eeq
where the  best fit   $ \rho \approx -3/2$. Again, as in the spherical CDT, all dimensionless constants appearing in eq. \eqref{LeffCTorus}, i.e.,  $\tilde  \Gamma, \tilde  \lambda$ and  $\tilde  \xi$ 
depend on the bare couplings of CDT ($k_0$ and $\Delta$). 
The kinetic term and the cosmological part of the potential term in equation \eqref{LeffCTorus} are consistent with the toroidal minisuperspace  action \eqref{CSMSVT}.\footnote{As already noted, the CDT effective action and the minisuperspace action agree up to overall sign, see discussion in Conclusions.} 
The additional potential term $\propto V_3^{-3/2}$, where we used the continuum physics notation,  does not have a classical counterpart and therefore  can be treated as a genuine quantum correction. At the moment it does not have a clear  interpretation.

\subsubsection{The effective action in other phases}

Using the transfer matrix method discussed above one can as well determine the form of the effective action in other phases of CDT. In particular the effective transfer matrix data measured in phase $A$ are well fitted by \cite{EffactionJGS}
\beql{MTAphase}
\langle{n|{ M}^{(A)}|m}\rangle ={\cal N}_0 \exp\big( \tilde \lambda(n+m) -\tilde  \xi (n^\rho + m^\rho)  \big) , \quad \rho\approx 0.5 \ ,
\eeq
yielding an effective action without a kinetic term.
One can check that indeed when approaching the $A-C_{dS}$ transition from inside the semiclassical phase $C_{dS}$ the measured value of $\tilde \Gamma \to \infty $, causing the kinetic term of the effective actions \eqref{eq:Sdiscrete} and \eqref{LeffCTorus} disappear at the transition point. Lack of the kinetic term inside phase $A$ can be therefore interpreted as the limit of the effective Newton’s constant $G_{eff} \to \infty$, leading to a causal disconnection of different spatial slices.

The situation is more complicated in phase $C_b$ where the transfer matrix is well parametrized by \cite{EffactionJGS}
\beql{TMCBphase}
    \langle{n|{M}^{(C_b)}|m}\rangle  =
    {\cal N}_0
    \left[ \exp \left( -2 \frac{\Big((m-n) -\big[ c_0(n+m-s^b) \big]_+\Big)^2}{\tilde \Gamma(n+m-2 n_0)}\right) + \right.
\eeq
$$
    \left.
    +\exp \left( -2\frac{\Big((m-n) +\big[ c_0(n+m-s^b) \big]_+\Big)^2}{\tilde \Gamma(n+m-2 n_0)}\right) 
    \right]\exp(-V[n+m]) \ ,
$$
where: $[...]_+ \equiv \max(... , 0)$, and the potential part $V[n+m]$ was not determined precisely. 
As can be seen from equation \eqref{TMCBphase} the kinetic term of the effective transfer matrix bifurcates for the spatial volume $n+m > s^b$ (the bifurcation point). Hence the other name of the phase $C_b$, also called the {\it bifurcation} phase. When approaching the $C_b-C_{dS}$ phase transition from inside the bifurcation phase $C_b$ one observes $s_b \to \infty$ and $c_0\to 0$, resulting in the effective action change at the  transition point. Accordingly, in phase $C_{dS}$ one recovers the  standard kinetic term of the minisuperspace action. It can be shown that the bifurcation of the kinetic term in phase $C_b$ is related to the appearance of high order  vertices in generic triangulations observed  in that phase. Such vertices are surrounded by large volume clusters, causing that the volume density is manifestly non-homogeneous, and therefore the minisuperspace description ceases to be valid inside the bifurcation phase $C_b$.

\subsection{Summary of the semiclassical limit}

We have shown, that a quantum theory of gravity defined by the four-dimensional Causal Dynamical Triangulations correctly predicts a semiclassical universe with positive cosmological constant, which on large scales is
extended and four-dimensional, and whose shape is compatible with a
cosmological solution of classical General Relativity. Remarkably, these properties have been derived from first principles 
in  the fully background independent and non-perturbative lattice QFT approach, in which the ground state   geometry  emerges dynamically from quantum fluctuations.
Inside phase $C_{dS}$ the fluctuations of the scale factor are very well described by a simple minisuperspace action. In the thermodynamical limit (when the lattice size $N_4\to \infty$) the (relative) amplitude of quantum fluctuations vanishes and one recovers a fully classical trajectory of the scale factor, matching our current understanding of the physics of the very early Universe, at least  under the assumption  of a spatially homogeneous and isotropic spacetime.

\section{Search for a continuum limit}

As already mentioned in the introduction, lattice field theories are well suited to study 
 non-perturbative aspects   of quantum field theories,
such as a (potentially non-perturbatively renormalizable) QFT describing quantum gravity. In the lattice formulation one regularizes a QFT in question by introducing a lattice of size $N$, where the lattice spacing $a$ plays a role of a UV cutoff $\sim 1/a$. Then, in order to define the continuum limit, one has to remove the cutoff by taking  $a \to 0$ and $N\to \infty$ in a controlled way, such  that the continuum physics quantities of a given QFT remain unchanged. This is related to the renormalization group flow, which describes how the dimensionless bare lattice  couplings should be adjusted when approaching the continuum limit. In particular, the QFT is usually characterized by some physical  length scale  which can be related to a correlation length $\xi$. While taking the continuum limit $a \to 0$, the correlation length $\xi_{a}$   measured in terms of (decreasing) lattice units $a$ must  diverge, i.e.,  $\xi_a \to \infty$ when $a \to 0$, such that the  physical correlation length  $\xi = \xi_a  a $ is kept fixed. The diverging correlation length is characteristic for a higher-order phase transition. 
 Therefore the presence of the higher-order  transitions in a lattice theory is a {\it sine qua non} condition of the existence of  the continuum limit and the related renormalization group fixed point(s).
 We thus have a space of (dimensionless)
 bare coupling constants associated with the lattice theory and we want to
locate regions in this space where there are higher-order phase transitions.
 In this section we will discuss how the order of a phase transition can by measured using numerical MC  methods and we will provide evidence that (at least some of) the CDT phase transitions  are indeed the higher order transitions. Then we will try to define the CDT renormalization group flow, following the paths of "constant physics" in the direction of decreasing lattice spacing. 

\subsection{Phase transitions}

Phase transitions can typically be captured by analyzing some macroscopic properties of a given 
system  and tracking their changes when varying  coupling
constants of a theory in question. The notion of an order of a
phase transition was introduced by Ehrenfest, who proposed to characterize phase transitions using derivatives of  thermodynamical potentials, such as free energy,  measured at the transition point. According to the Ehrenfest classification, if all first $(n-1)$ derivatives of the free energy are continuous and the $n$-th order derivative diverges at the transition point then one has the $n$-th order phase transition.  This picture was refined by Landau  and Ginzburg who  introduced the notion of  an order parameter ($OP$).\footnote{One should note that in some recent models of solid state physics the Ehrenfest-Landau-Ginzburg characterization can fail, when a phase transition cannot be characterized by a local order parameter, but rather by long-range entanglement, called topological order. Nevertheless in the following we will stick to the proposed classification of phase transitions.}   The order parameter is related to the first-order derivative of the free energy and it captures the symmetry difference between the two different phases. Since then the classification of phase transitions shifted towards distinguishing between two types of phase transitions: first-order, which has a divergent first-order derivative of the thermodynamic potential,  and higher-order (also called continuous), where the first-order derivative is continuous but second- or higher-order derivatives diverge. Therefore, in the case of a first-order transition an order parameter is discontinuous at the transition point and one observes a $\delta$-like singularity of susceptibility related to a second-order derivative of a thermodynamical potential. The behaviour is different for a higher-order phase transition, where  an order parameter  is continuous   but it is non-analytic at the transition point. As a consequence the second- or higher-order derivatives diverge in a specific way. Furthermore, in the
above classification, there is a relation between the order of a transition and the
correlation length. For a first-order transition one typically has finite correlation
length, while divergent correlation length signals a continuous phase transition. 
As will be discussed below, the distinction between orders of phase transitions using lattice methods is non-trivial. This is caused by the fact that in a lattice formulation one deals with a large but finite number of degrees of freedom and thus both the thermodynamical potentials and all their derivatives are finite. Therefore, instead of observing  true phase transitions one in fact deals with {\it pseudo phase transitions} (cross-overs), causing much  smoother transition signals. It is only in the thermodynamical  limit, i.e., when the lattice size $N\to \infty$, that one recovers the true phase transitions.  It is therefore necessary to analyze  finite-size effects and discuss how the  lattice results can be interpolated to the   infinite  volume (lattice size) limit.

\subsubsection{Ising model example}

As a toy model example let us discuss the Ising model of a ferromagnet with the  nearest neighbours interaction on a ($d$-dimensional) square lattice of size $N$ (denoting the number of vertices where  spins are placed),
described by a partition function
\begin{equation}
    {Z}(\beta,h) = \sum_{\{s_i\}} \exp \left({ \beta \sum_{i\leftrightarrow j} s_i s_j + h \sum_i s_i}\right), \quad s_i=\pm 1,
\end{equation}
where $\{s_i\}$ in the first sum denotes all possible spin
configurations. In the model there are two (dimensionless) coupling constants $\beta$ and $h$, related respectively to the inverse temperature and the external magnetic field.\footnote{We used a (non-standard) convention in which $h$ is the inverse temperature $\times$ the external field.} In the thermodynamical limit, for $d\geq 2$ dimensions  and without external field ($h=0$)  one has a second-order (continuous) phase transition between the ordered and the disordered phase occurring at the critical (inverse) temperature $\beta_c$.\footnote{In $d=2$ dimensions $\beta_c= \frac{1}{2} \ln (1+ \sqrt 2) \approx 0.44$} The Ising model with external field can be also used to describe a first-order phase transition between positive and negative magnetization occurring for $\beta > \beta_c$ (low temperature) at $h_c=0$.

\noindent \textbf{Higher-order transition}

Let us start with a second-order phase transition in the absence of the external field ($h=0$). The macroscopic properties of the system can be traced by magnetization (per spin) 
\beql{magnetperspin}
    m=\frac{1}{N}\sum_i s_i.
\eeq
It plays a role of  an order parameter distinguishing between the low temperature (high $\beta$) ordered phase, where the system is magnetized ($\langle m  \rangle \neq 0$), and the high temperature (low $\beta$) disordered phase, where the magnetization vanishes ($\langle m \rangle \approx 0$). In the disordered phase the spins are distributed randomly and one recovers a (spin-inversion) $Z_2$ symmetry of the Hamiltonian (with $h=0$), while in the ordered phase the spins prefer to align together and therefore spontaneously break the $Z_2$ symmetry. Accordingly, the order parameter captures the difference between symmetries of generic spin configurations observed in the two phases.\footnote{Note that formally  the $Z_2$ symmetry is present also in the ordered phase as in any finite temperature there is a non-zero probability of tunnelling between the positive and the negative magnetization which are equally probable. This can be  observed for small lattices. Therefore  one often uses $|m|$ instead of $m$ as an order parameter, as $\langle|m|\rangle$ is manifestly non-zero in the ordered phase. The tunnelling probability decays exponentially with $N$ thus in practice for large lattices the tunnelling never happens in the MC simulation time, and one can as well use $\langle m \rangle$.}
The average magnetization
\begin{equation}
    \langle m \rangle =  \frac{1}{N}  \Big\langle {\sum_i s_i} \Big\rangle = \frac{1}{N} \frac{1}{{Z}}\sum_{\{s_i\}} \sum_i s_i \exp \left({ \beta \sum_{i\leftrightarrow j} s_i s_j + h \sum_i s_i}\right)=  \frac{1}{N} \frac{\partial \ln {Z}}{\partial h},
\end{equation}
 is related to a  first-order derivative of the free energy $F=-\frac{1}{\beta}\ln {Z}$. 

One can as well compute the (magnetic) susceptibility $\chi$ which is the first-order derivative of the (average) magnetization and thus the second-order derivative of the free energy

\begin{equation}
    \chi = \frac{\partial \langle {m} \rangle }{\partial h} = \frac{1}{N} \frac{\partial^2 \ln {Z}}{\partial h^2}.
\end{equation}
It follows  that
\begin{equation}
    \frac{\chi}{N} 
    =  \langle {m}^2 \rangle - \langle m \rangle^2,
\end{equation}
so the susceptibility $\chi$ is proportional to the magnetization variance. The susceptibility 
\begin{equation}
    \chi =\frac{1}{N} \sum_{i \,j} \big( \langle s_i s_j \rangle - \langle s_i \rangle \langle s_j \rangle \big) = \frac{1}{N} \sum_{i \, j } G^{(2)}_c(i,j) = \sum_{j } G^{(2)}_c(0,j)
\eeq
is also related to the (connected)  two-point  correlation function defined by
\beql{G2Ising}
    G^{(2)}_c(i,j) = \frac{\partial^2 \ln {Z}}{\partial h_i \partial h_j},
\eeq
where we have introduced a space-dependent external field ($h \to h_i$). Generically, away from the transition point the correlation function  decays exponentially
\beql{Gexp}
    G^{(2)}_c(i,j) \sim e^{-|x_{ij}|/\xi},
\end{equation}
where $|x_{ij}|$ is the distance between two lattice points $i$ and $j$, and $\xi$ is the correlation length, 
which depends on the parameters of the system, i.e., $\xi=\xi(N,\beta,h)$. Typically the correlation length is of order of a few lattice units $a$.
Interesting physics arises when $\xi \to \infty$. This is achieved in the thermodynamical limit $N\to \infty$ by fine-tuning the coupling constant(s) to the critical value(s), i.e., for the Ising model $\beta \to \beta_c$. In the neighborhood of a continuous phase transition the exponential fall-off in equation \eqref{Gexp} vanishes and the correlator starts to behave according to the power law
\begin{equation}
      G^{(2)}_c(i,j) \sim \frac{1}{|x_{ij}|^{d-2+\eta}}
\end{equation}
where $\eta$ is a critical exponent. As a result correlations extend to macroscopic distances $|x_{ij}|\gg a$ and the system starts to be insensitive to the short-distance details of the lattice.  Therefore, in the vicinity of the higher-order critical point, one can possibly obtain a continuum limit of a theory microscopically defined on a lattice. As microscopic details of the lattice become unimportant  there will be many theories with a different microscopic  formulation but the same properties in the continuum limit, such a phenomenon is called {\it universality} and it is mostly governed by  the same symmetries and dimensionality of different systems. 
The  (universal) non-analytic  behaviour near a continuous phase transition point is captured by scaling relations. In the thermodynamical limit $N\to \infty$ one has in the transition region
\beql{ksiscale}
    \xi \sim |\beta_c-\beta|^{-\nu} \ ,
\end{equation}
\begin{equation}
    \langle m \rangle \sim |\beta_c-\beta|^{\alpha} \ ,
\end{equation}
\beql{suscscale}
    \chi \sim |\beta_c-\beta|^{-\gamma} \ ,
\end{equation}
where $\nu$, $\alpha$ and $\gamma$ are critical exponents.\footnote{The critical exponent $\alpha$ is usually denoted $\beta$ but here we used a different notation to distinguish it from the (inverse) temperature. For the two-dimensional Ising model one has $\nu=1$, $\alpha=1/8$, $\gamma=7/4$.} For a finite lattice of size $N$ the pseudo-critical point $\beta_c(N)$ is shifted relative to the true critical point $\beta_c=\beta_c(\infty)$. Also the largest  length scale available on a finite lattice is of the order of the linear extension of the lattice  $L=N^{1/d}$ and thus instead of the infinite correlation length at the phase transition one has $\xi(N)\sim N^{1/d}$. Using equation \eqref{ksiscale} one obtains
\beql{scaleksinu}
    |\beta_c - \beta_c(N)|^{-\nu} \sim N^{1/d} ,
\eeq
leading to the following finite-size scaling relation of the pseudo-critical  point
\beql{scalepos}
    \beta_c(N)=\beta_c - \text{const} \times N^{-\tilde \nu}\ , \quad \tilde \nu = \frac{1}{\nu d}.
\eeq
As suggested by equation \eqref{suscscale}, for  fixed lattice volume $N$ one can find  a pseudo-critical point $\beta_c(N)$ by observing a peak in susceptibility.
The scaling exponent $\tilde \nu$ can be then computed numerically by performing a series of  MC measurements for larger and larger lattices  and fitting the scaling relation~\eqref{scalepos}.  For a higher-order (continuous) transition one expects $\tilde \nu < 1$, e.g., for the two-dimensional Ising model one has $\tilde \nu = 1/2$. 

Similarly, using equation \eqref{scaleksinu} one obtains the following finite size scaling relations 
\beql{scaleM}
    \langle m \rangle  \sim N^{-\tilde \alpha} \ , \quad \tilde \alpha = \alpha \tilde \nu,
\end{equation}
\beql{scalechi}
    \chi \sim  N^{\tilde \gamma} \ , \quad \tilde \gamma = \gamma \tilde \nu.
\end{equation}
The scaling exponents $\tilde \alpha$ and $\tilde \gamma$ can again  be computed numerically by measuring the values of  the (average) magnetization and susceptibility   at the pseudo-critical points $\beta_c(N)$ and then fitting equations \eqref{scaleM} and \eqref{scalechi}, respectively. For higher order transitions one  expects $\tilde \alpha$ ,  $\tilde \gamma < 1$.

\noindent \textbf{First-order transition}

The Ising model that we have been investigating also exhibits a  phase transition between  positive and negative magnetization which can be achieved for $\beta > \beta_c$ (low temperature) through the introduction of an external magnetic field $h$. By the symmetry it is expected that the transition occurs at $h_c=0$, where the (average) magnetization $\langle m \rangle$ jumps from positive to negative values. The jump in magnetization means that there is a discontinuity of the first-order derivative of the free energy and thus the transition is first-order. Consequently, one can also expect to observe a $\delta$-like singularity in the magnetic susceptibility $\chi$, which is a first-order derivative of the magnetization and the second-order derivative of the free energy. 

Conversely to the higher-order transition case,  the correlation length  stays finite and there  are no long-range collective fluctuations in the system.  As a result the rearrangement of spin configurations from positive to negative magnetization is difficult. In the thermodynamical limit $N\to \infty$ one observes a hysteresis which causes the phase change to occur at  points slightly displaced from $h_c=0$. The hysteresis occurs due to metastable states of positive and negative magnetization  coexisting at the critical point.  
In numerical simulations, the hysteresis  is visible only for large enough lattices, as for large $N$ the MC simulation time is usually shorter than the MC autocorrelation time  at the transition point, which is large and grows exponentially with (some power of) $N$. For smaller lattices, where the autocorrelation time is much shorter than the MC simulation time, one can observe   back and forth tunnelling between the two metastable states in the transition region ($h\approx 0$). This causes that the transition is smoothed and  thermodynamical  quantities become continuous and "rounded" for finite  $N$, e.g., instead of the jump in magnetization one observes a $\tanh$-like behaviour of $\langle m \rangle$, similarly instead of $\delta$-like behaviour of susceptibility there is a smooth peak in $\chi$.
This shows that in practice the accurate estimation of jumps in the expectation value of an order parameter and singularities in its susceptibility at phase transitions which are only "weakly" first-order is particularly cumbersome and then even the distinction from second-order transitions may be a problem. Therefore again one has to  perform a careful finite-size scaling analysis in order to extrapolate from the smooth behaviour of a finite system towards jumps and singularities of $N\to \infty$.

Typically, in the case of a first-order  transition
the probability distribution of an order parameter in the transition region   can be described by a sum of two shifted Gaussians, centered at values characteristic for the two metastable states coexisting at the transition point.  In particular, in the Ising model the behaviour of magnetization $m$ is well approximated by the following  probability distribution \cite{Binder}\footnote{For the Ising model the double Gaussian approximation  can be derived using Landau-Ginzburg theory formulated in therms of  free energy.  It is not accurate in the region between the two peaks, since there $p(m)$ reflects interfacial contributions of spin clusters, but (for large enough $N$) probability density in this region is very small so it contributes  only negligible corrections  to $\langle m \rangle$ and $\chi$.}
\beql{Ising2Gauss}
    p(m)= {\cal N} e^{h m N} \left[  \exp(-\frac{(m-\bar m)^2}{2 \bar \chi / N })+ \exp(-\frac{(m+\bar m)^2}{2 \bar \chi / N })\right]
\eeq
$$
    = \tilde {\cal N} \left[e^{h \bar m N}  \exp(-\frac{(m-\bar m - h \bar \chi)^2}{2 \bar \chi / N })+ e^{-h \bar m N} \exp(-\frac{(m+\bar m- h \bar \chi)^2}{2 \bar \chi/ N })\right],
$$
where $\cal N$, $\tilde{\cal N}$ are normalization factors.
For $h=0$ one recovers two equally probable metastable states of magnetization centered at $\pm \bar m$ and the amplitude of fluctuations  within each state is governed by $\bar \chi$ (note that $\bar \chi$ is not the total susceptibility but rather a susceptibility within each state).\footnote{In other models the positions of states don't  have to be symmetric w.r.t. zero and also the widths of the two Gaussians can differ.} The width of the two Gaussians decreases with $N$ and in the limit $N\to \infty$ one obtains two shifted $\delta$-functions. It is also visible that the non-zero external field $h$ shifts the position of the two Gausians in the direction of $h$ and the shift is proportional to $h$. At the same time the  relative probability of observing each state changes with $h$, such that for $h>0$ the  peak  at positive $m$ becomes  exponentially larger than the  peak at negative $m$, while for $h<0$  the opposite happens and thus negative magnetization prevails, and the effect is stronger for larger $N$.

It is a simple task to compute $\langle m \rangle$ and $\chi$ using probability distribution \eqref{Ising2Gauss}. One obtains
\beql{IsingM}
    \langle m \rangle = \bar \chi h + \bar m \tanh(h \bar m N),
\eeq
\beql{IsingChi}
    \chi = \frac{\partial \langle m \rangle}{\partial h} = \bar \chi + \bar m^2 N / \cosh(h \bar m N) .
\eeq
This  explains the smoothed $\tanh$-like behaviour of magnetization (instead of a jump) as well as a smooth peak of height proportional to $N$ in susceptibility (instead of a $\delta$-like function)  observed for finite $N$. 
Away from the critical point, i.e., for $|h|\gg |\bar m N|^{-1} $ one recovers the expected "bulk" behaviour 
\begin{equation}
    \langle m \rangle \approx   \bar \chi h \pm \bar m , \quad \chi \approx  \bar \chi,
\end{equation}
whereas in the transition region, i.e., for $|h|\ll |\bar m N|^{-1} $,  one has
\beql{scale1order0}
    \langle m \rangle \approx   h \bar m^2 N , \quad \chi \approx   \bar m^2 N.
\end{equation}
It follows that at the first-order critical point one obtains the following universal finite-size scaling relations\footnote{For the Ising model one formally has $\langle m \rangle$ = 0 at $h_c=0$.}  
\beql{scale1order}
    \langle m \rangle \sim  N , \quad \chi \sim N.
\end{equation}
Note, that the only scaling which enters the above equations is related to the total volume  of the system (lattice size $N$). This is 
different than the scaling observed  for higher-order transitions, see equations \eqref{scaleM} and \eqref{scalechi}, where one has non-trivial scaling exponents.
In the Ising model discussed here, due to the spin-inversion symmetry at $h=0$, there will be no shift of the pseudo-critical point with $N$ as always $h_c(N)=0$. Nevertheless, in other models which exhibit first-order transitions one can also expect a finite-size scaling of the pseudo-critical point, similar to equation \eqref{scalepos}, where again the scaling will be related to the total volume $N$, i.e., one has a trivial scaling exponent $\tilde \nu =1$, being  the only power of $N$ which comes into play for the first-order transition.

\subsubsection{CDT phase transitions}

As discussed above, phase transitions
belonging to the same universality class will show the same type of finite-size
scaling properties, which manifest  by universal values of scaling exponents and  in principle can be used to distinguish between first- and higher-order transitions. 
The Ising model example shows that a practical distinction between the two types of phase transitions   using lattice methods is non-trivial and requires a careful analysis of  finite-size effects. In both cases  pseudo-critical   points can be found by tracking an order parameter ${\cal O}$, used to distinguish the two different phases, and observing a peak in  its susceptibility
\begin{equation}
    \chi_{{\cal O}}\equiv \bar N_4 \Big(\langle {\cal O}^2 \rangle -\langle {\cal O} \rangle^2\Big),
\end{equation}
where $\bar N_4$ denotes the lattice size, i.e., the (constant) target four-volume fixed in the MC simulations of CDT.\footnote{In CDT one often performs MC simulations by fixing $\bar N_4\equiv \bar N_4^{(4,1)}$, i.e., the total number of $T^{(4,1)}$ and $T^{(1,4)}$ simplices, instead of true $\bar N_4=\bar N_4^{(4,1)}+ \bar N_4^{(3,2)}$, also including $T^{(3,2)}$ and $T^{(2,3)}$ simplices.} 
In the MC simulations, for each lattice size $N$, one typically changes values of one coupling constant  (below denoted $\cal C$, in the case of CDT: ${\cal C}\equiv k_0$ or $\Delta$, respectively) while keeping other coupling(s) fixed and one can therefore measure the finite-size scaling of the pseudo-critical point ${\cal C}_c(N)$
\beql{scaleposCDT}
    {\cal C}_c(\bar N_4)={\cal C}_c - \text{const} \times \bar N_4^{-\tilde \nu},
\eeq
where the critical exponent $\tilde \nu=1$ indicates a first-order transition while $\tilde \nu < 1$ a higher order transition. One could  look as well at the average value of an order parameter $\langle {\cal O} \rangle$ or its susceptibility $\chi_{{\cal O}}$ measured at the pseudo-critical points ${\cal C}_c(N)$ and use finite-size scaling relations
\beql{scaleOPDT}
    \langle {\cal O} \rangle  \sim \bar N_4^{-\tilde \alpha},
\end{equation}
\beql{scalechiCDT}
    \chi_{{\cal O}} \sim  \bar N_4^{\tilde \gamma},
\end{equation}
where again $-\tilde \alpha =1$ and $\tilde \gamma =1$ point to a first-order transition while other values of the critical exponents indicate a higher-order transition.
In practice the relations \eqref{scaleOPDT} and \eqref{scalechiCDT} are rarely used in CDT as the observed  phase transitions are quite "narrow" and
the resolution of the numerical data is insufficient to measure the average value  of an order parameter at the transition point or the susceptibility peak value with good enough accuracy. 

Another possibility of distinguishing between  first- and  higher-order phase transitions comes from observing two different metastable states present for the first-order transition and absent for the higher-order transition. For example,  one can  analyze empirical distributions 
 (histograms) of order parameters measured at  pseudo-critical points. 
As  explained for the Ising model, the two metastable states cause double peaks in the  probability distributions. If a phase transition is "strongly" first-order one can also observe hysteresis in the transition region. However, as CDT results show,  one should again very carefully   analyze finite-size effects, as it can happen that the hysteresis  (or the two peaks separation) shrinks to zero in the $N\to \infty$ limit.
The appearance of double peaks in the histograms is well quantified by the, so-called, Binder cumulant, related to the 4-th moment of the  probability distribution\footnote{Note that here, and in general in the CDT phase transitions studies, we use a definition of the Binder cumulant which is shifted (by a  $-2/3$ constant) versus the original Binder's definition  \cite{BinderPhysRevLett}: $B_{\cal O}=1- \frac{1}{3} \frac{\langle {\cal O}^{4} \rangle}{\langle {\cal O}^2\rangle^{2}}$.}
\begin{equation}\label{binder}
    B_{{\cal O}}\equiv\frac{1}{3}\left(1-\frac{\langle {\cal O}^{4} \rangle}{\langle {\cal O}^2\rangle^{2}}\right).
\end{equation}
For a first-order transition, where the  histograms   have two shifted peaks,  $B_{{\cal O}}$ measured at the critical point will move away from zero when $N\to \infty$ as the two peaks become more and more apparent (the probability distribution  approaches two shifted $\delta$-functions). If instead the transition is higher-order, then the histograms have only one peak and $B_{{\cal O}}$  should tend to zero for $N\to\infty$. Thus a deviation of $B_{{\cal O}}$ (measured at the critical point) from zero in the limit $N\to \infty$ can indicate a first-order transition while a value close to zero suggests a higher-order transition.
In Table \ref{Table1} we summarize the characteristics of both types of phase-transitions which were used in CDT to distinguish between the first- and the higher-order transitions. 

\begin{table}
\centering
\caption{ Characteristics of the first- and the higher-order phase transitions.}
\resizebox{\textwidth}{!}{
    \begin{tabular} {|c|c|c|c|}
        \hline
        \bf{   OBSERVABLE}	& \bf{  First-order transition}	& \bf{ Higher-order transition} \\ \hline
        \hline
        Pseudo-critical point 	&  $\tilde \nu\approx 1$	&   $\tilde \nu <1 $ \\
        scaling according to \eqref{scaleposCDT}	&  	&   \\ \hline
        Susceptibility scaling 	&  $\tilde \gamma\approx 1$	&   $\tilde \gamma <1 $ \\
        according to \eqref{scalechiCDT}	&  	&   \\ \hline
        Hysteresis & YES & NO \\
        in the transition region & and remains when $\bar N_4\to\infty$ & or disappears when $\bar N_4\to\infty$\\ \hline
        Histograms measured at	&   double peaks	&  single peak or  \\ 
        pseudo-critical points	&   peak separation $\uparrow$ with $\bar N_4\to\infty$	&   peak separation $\downarrow$ with $\bar N_4 \to\infty$  \\ \hline
        Binder cumulant measured	&   $B_{{\cal O}} < 0 $	&   $B_{{\cal O}} \approx 0$ \\
        at pseudo-critical points &   when	$  \bar N_4\to\infty $ &   when $ \bar N_4\to\infty$ \\ \hline
    \end{tabular}
}
\label{Table1}
\end{table}

\begin{figure}[ht!]
\centering
    {\includegraphics[width=0.45\textwidth]{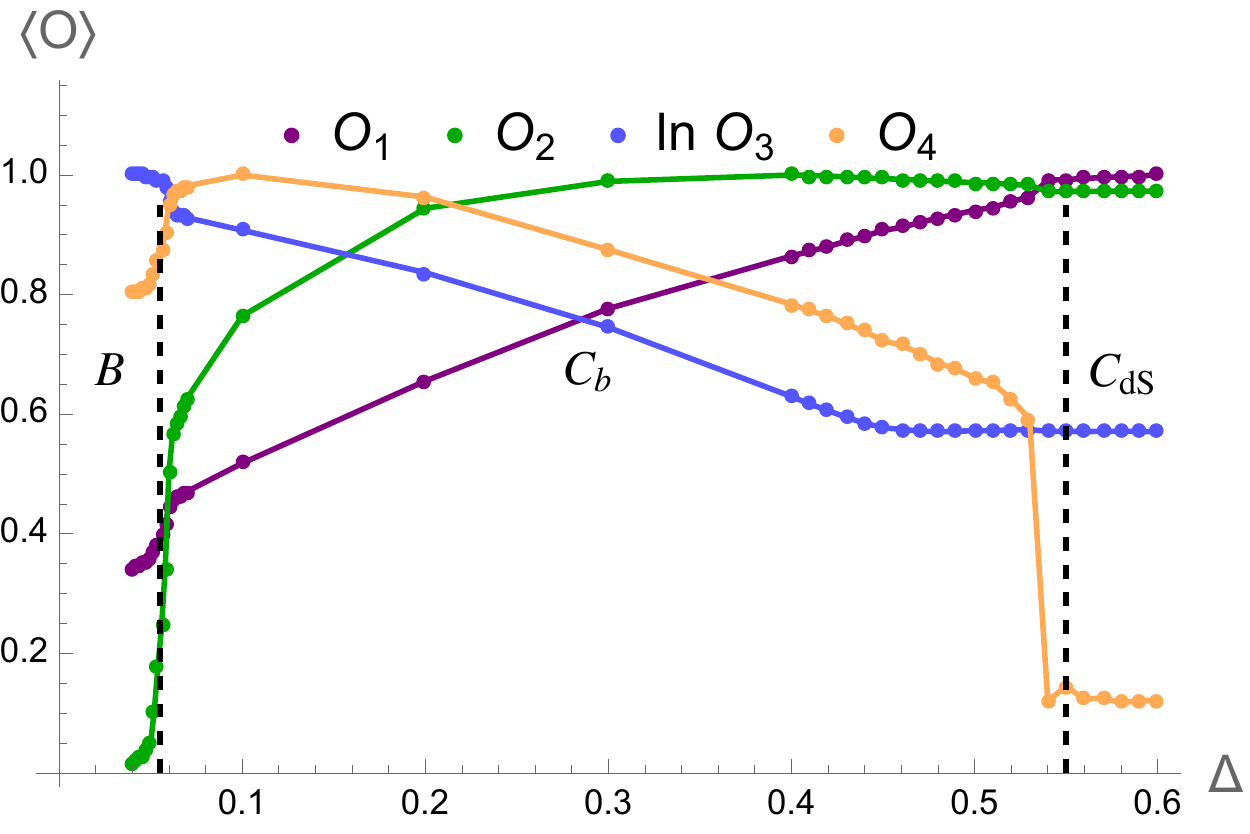}}
    {\includegraphics[width=0.45\textwidth]{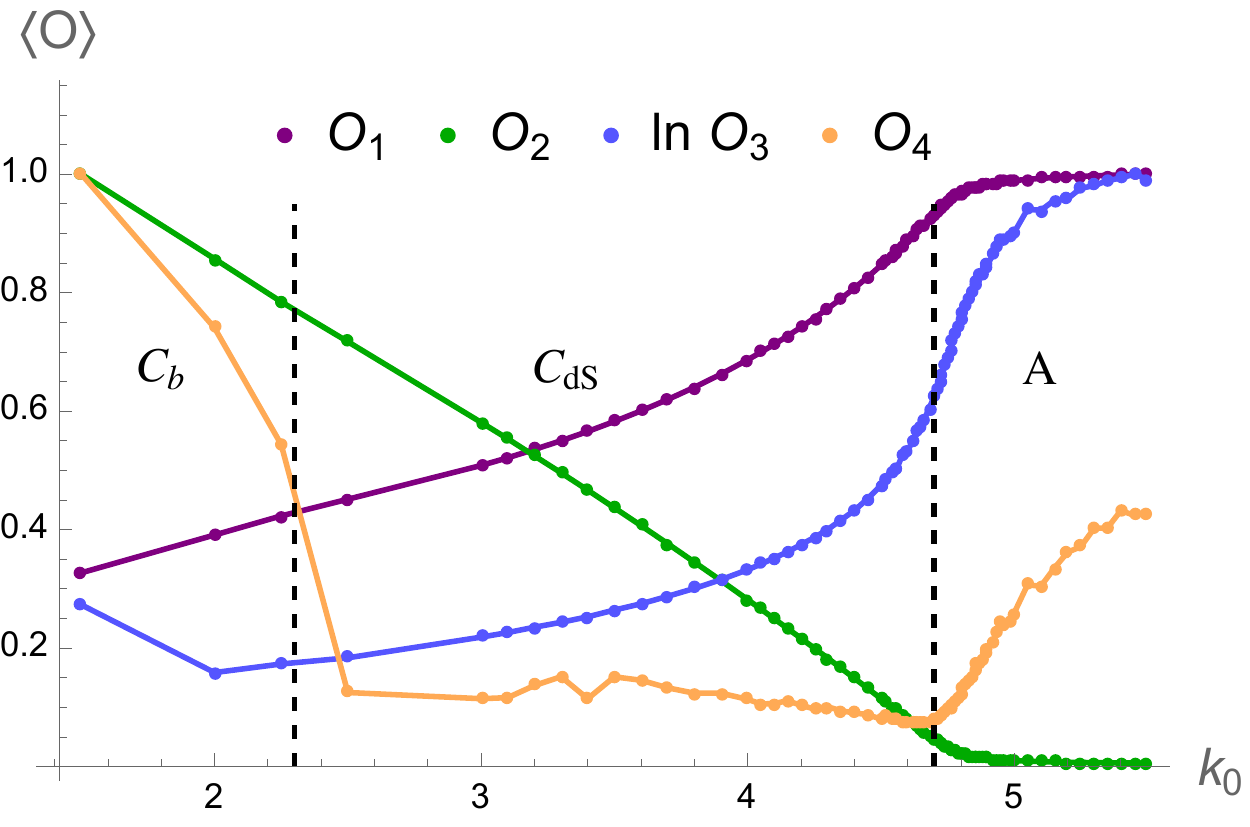}}
\caption{CDT order parameters, see equations \eqref{OP12} and \eqref{OP34}, measured in the toroidal CDT for fixed $k_0=1.5$ (left panel) and for fixed $\Delta=0.4$ (right panel). The order parameters were rescaled to fit into  range $[0,1]$. Phase transitions are indicated by dashed lines. }
\label{fig:OPs}
\end{figure}

Last but not least one should discuss what order parameters have been used in CDT. In the Ising model example the order parameter (the magnetization $m$)  was conjugate to the external field $h$ appearing in the Hamiltonian and playing a role of a coupling constant in the model. Similarly, in CDT one has three coupling constants: $k_0, \Delta$ and $k_4$ and one can thus use the conjugate quantities, i.e., the global numbers $N_0$, $N_4^{(4,1)}$ and $N_4^{(3,2)}$ characterizing a CDT triangulation and appearing in the Regge action. Usually in the MC simulations one fixes the total lattice volume $ N_4^{(4,1)}$. 
Consequently  one has the following  order parameters\footnote{Note that similarly to the Ising model, where one used an intensive magnetization per spin $m$, in CDT we also normalize by the  
lattice volume $N_4^{(4,1)}$.}
\beql{OP12}
    {\cal O}_1=\frac{N_0}{N_4^{(4,1)}} \quad , \quad {\cal O}_2=\frac{N_4^{(3,2)}}{N_4^{(4,1)}} .
\eeq
One can as well use more local information, not captured by the "global" order parameters defined above. For example,  even though the total lattice volume $N_4^{(4,1)}$ is fixed in the MC simulations, its distribution over spatial slices (the volume profile $N_3(k)$) is not  and the generic spatial volume distribution  can vary among different phases. Similarly, there is a global  relation between $N_0$ and $N_4^{(4,1)}+ N_4^{(3,2)}$, captured by an average coordination number of  a vertex\footnote{The coordination number is defined as a number of 4-simplices sharing a given vertex.}, but the typical coordination numbers of individual vertices can be very much different between generic triangulations observed in different phases. Therefore one can define two more order parameters
\beql{OP34}
    {\cal O}_3 = \frac{1}{N_{4}^{(4,1)}} \sum_{k}
\Big(N_3(k)-N_3(k+1)\Big)^2 \ \ , \ \ {\cal O}_4 = \frac{1}{N_{4}^{(4,1)}} \argmax_{v} \big(c(v)\big), 
\end{equation}
where $v$ denotes a set of all vertices in a  triangulation and $c(v)$ is the coordination number. 
The behaviour of all four  order parameters in various CDT phases is presented in Fig. \ref{fig:OPs}.
At a first glance it may seem a little bit surprising that in addition to the "typical" order parameters  ${\cal O}_1$ and ${\cal O}_2$, which are conjugate to the bare couplings of CDT appearing in the Regge action  \eqref{SR}, we have introduced new order parameters ${\cal O}_3$ and ${\cal O}_4$, not related to the bare action. We did it for pragmatic reasons, as the  new order parameters enable us to study some phase transitions, like the $C_b-C_{dS}$ transition, which are not visible using the "typical" order parameters. 
Nevertheless one could say that the same happens in
 the "pure" (meaning without the external field, i.e., for $h=0$)  Ising model, which  undergoes a higher-order phase transition at $\beta=\beta_c$. In order to define the expectation value and susceptibility of an order parameter (magnetization) in terms of derivatives of the free energy one has to "enlarge" the Ising model by introducing a new coupling $h$, related to the external field. Similarly, in the case of CDT one can "enlarge" the model by adding non-trivial couplings to local degrees of freedom, not appearing in the bare Regge action of CDT \eqref{SR}. For example, ${\cal O}_4$ has exactly such interpretation as it can be related to a  non-trivial measure term (dependent on local order of vertices) introduced in some  lattice   models of quantum gravity \cite{chapterEDT}. It turns out that indeed the order parameter ${\cal O}_4$ is very well suited  to detect the $C_b-C_{dS}$ transition related to the appearance of high-order vertices, characteristic for phase $C_b$ and not present in phase $C_{dS}$, see Fig. \ref{fig:OPs}.
 
 Using the  order parameters ${\cal O}_1-{\cal O}_4$ one can measure phase transition characteristics described above (see Table \ref{Table1}) and thus in principle distinguish between the first- and the higher-order transitions. In practice, the distinction  is not always easy as phase transition signals are  mixed for most CDT  transitions. Therefore, we take a cautious approach, i.e., even in a case where most characteristics  point to a higher-order transition but there are some signals of a first-order transition  we classify the transition in question as the first-order transition. Using these rules one has the following:
 in both the spherical and the toroidal CDT, the $A-C_{dS}$ transition was found to be a first-order transition \cite{SamoLong, CriticalPhenomena}, while the $B-C_b$ transition is a higher-order transition \cite{SamoLong,Czelusta}. As regards other CDT phase transitions the picture is more complicated.  The $C_b-C_{dS}$ transition was found to be a higher-order transition in the spherical CDT \cite{ExploringCoumbe}, but a first-order transition in the toroidal CDT \cite{PhasestructureTorus}. The $B-C_{dS}$ transition and the $A-B$ transition was so far studied only in the toroidal model, where both transitions were found to be first-order \cite{TopologyInduced}. The above results are summarized in Table \ref{Table2}.
 The order of the transitions is also depicted in Fig. \ref{fig:phasediagram}, where blue solid lines denote  first-order transitions, while red solid lines are higher-order transitions.
 
\begin{table}
\centering
\caption{ Summary of CDT phase transitions.}
    \begin{tabular}{|>{\centering}m{0.32\textwidth}|>{\centering}m{0.32\textwidth}|>{\centering\arraybackslash}m{0.32\textwidth}|}
        \hline
        \bf{  PHASE TRANSITION}	& \bf{  Spherical CDT}	& \bf{ Toroidal CDT}  \\ 
        \hline
        \hline
        $A-C_{dS}$	&  first-order	&  first-order \\
        \hline
        $B-C_{dS}$	&  ?	&  first-order \\
        \hline
        $C_b-C_{dS}$	&  \bf{higher-order}	&  first-order \\
        \hline
        $B-C_b$	&\bf{higher-order}	&  \bf{higher-order} \\
        \hline
        $A-B$	& ? &  first-order  \\
        \hline
    \end{tabular}
\label{Table2}
\end{table}

Let us stress that the CDT phase  transitions are non-standard from the point of view of ordinary phase transitions, since they involve changes in spacetime geometry and 
the related  effective  topology, rather than changes in field configurations on a fixed spacetime, like, e.g., in the Ising model. First, let us clarify what we mean by {\it effective topology}. In order to understand that let us consider spacetime
configurations observed in the de Sitter phase $C_{dS}$ in the spherical CDT, i.e., when the fixed spatial topology is $S^3$ and (due to time-periodic boundary conditions) the total topology of the triangulations is that of $S^1\times S^3$. By construction, any MC update of a triangulation cannot change the
imposed total topology. However, a generic configuration in phase $C_{dS}$ will
effectively look like an $S^4$-triangulation where the north-pole and the south-pole of the four-sphere
are connected by a thin "stalk" of cut-off size, cf. average spatial volume profiles in Fig. \ref{fig:vol_prof}.  The "stalk" is necessary to preserve the imposed total topology of the triangulations, but if only it was allowed by the MC algorithm it would completely disappear, yielding the change of  the topology from  $S^1 \times S^3$ to $S^4$ (which we call the effective topology). 
It is thus visible that not only the effective spacetime dimension but also the effective spacetime topology  can be emergent  concepts on the quantum level.

Now an interesting pattern emerges linking the order of  the CDT phase transitions with the effective topology of spacetimes observed in the adjacent phases. As  discussed above, the effective topology of phase $C_{dS}$ in the spherical CDT is that of $S^4$.
 The same effective topology is also observed (both in the spherical and the toroidal CDT) in the bifurcation phase $C_b$. At the same time  the spherical CDT $C_b-C_{dS}$ transition  is a higher-order transition. Also the effective topology of phase $B$ is similar  and the $B-C_b$ transition is a higher-order transition (both in the spherical and the toroidal case).
All other
phase transitions observed so far in CDT are first-order transitions and they are related to a change of the
effective topology. In the case of toroidal CDT, the $B-C_{dS}$ transition and the $C_b-C_{dS}$ transition involve
a change of the effective topology from $S^4$ (observed in phases $B$ and $C_b$) to $T^4$ (in the toroidal version of phase $C_{dS}$).  The effective topology of phase $A$ may be characterized as the topology of a disjoint union of spatial geometries, and thus it is different from the effective topologies encountered in other CDT phases, and again the $A-B$ and the $A-C_{dS}$ transitions are first-order transitions.
The above  results  lead to the following conjecture: 

{\it  phase transitions which involve a change in effective topology will be first-order transitions.}

 At the same time all CDT  transitions which don't involve the effective topology change are most likely higher-order transitions.

\subsection{RG flow in CDT and the perspective continuum limit}\label{sec:RGflow}

 Our main interest is to find a higher-order phase transition where one may be able to define a continuum limit of CDT. Since only the  phase $C_{dS}$ seems to offer acceptable infrared configurations, it is natural to look for the higher-order transition lines at the border of that phase. 
From the results reported above, we cannot use the toroidal CDT in such studies, since all such  transition lines are first-order.\footnote{There is still a chance that the triple points, i.e., the points where the three transition lines meet, see Fig. \ref{fig:phasediagram},  may potentially be higher-order critical points, as it is quite common to observe a higher-order critical point in the end of a first-order transition line. But this is hard to  check  numerically.}
 Therefore we will focus  on the  spherical CDT, where one observes that the $C_b-C_{dS}$ transition line is higher-order. According to our conjecture, it is also likely that the spherical $B-C_{dS}$ transition (whose order has not been determined yet)  is a higher-order transition, as it separates phases of the same effective topology. It also potentially makes the $C_b-B-C_{dS}$ triple point an interesting candidate for a UV-fixed point. At least it would then be a critical point where three higher-order transition lines meet.

Having  located the higher-order phase transition line(s)  we now want to understand how to approach the critical point(s) from inside the physically interesting phase $C_{dS}$. The general idea is quite simple: we would like to adjust 
the dimensionless bare couplings  of CDT  in such a way that the RG  trajectories in the $(k_0,\Delta)$ coupling constant space, i.e.,  the lines of "constant physics"  defined by some renormalized continuum observables,  flow  in the direction of decreasing lattice spacing $a$ and hopefully lead to the higher-order critical point, 
implying that the correlation length $\xi_a \to \infty$ and the lattice spacing $a\to 0$. Then such a critical point will correspond to a UV fixed point in which one obtains a continuum limit of the CDT lattice theory.  The absence of a UV fixed point would be signaled by the fact that there is not any RG flow line ending at a higher-order transition point. In that case one cannot shrink the lattice spacing to zero and obtain a continuum limit.
In  CDT it is convenient to analyze the RG flow of the bare couplings  under the additional assumption that the (average) physical volume of spacetime  is fixed and finite along each RG flow trajectory
\beql{ren1}
    V_4 \propto N_4 a^4 = \text{const.}
\eeq
 Accordingly, one has to increase lattice volume $N_4$ in an appropriate way when the lattice spacing $a$ is decreased, and the thermodynamical limit $N_4\to \infty$ will coincide with the continuum limit $a \to 0$, if it exists.

The key question remains: how to define what constitutes a line of "constant physics"? In the following we will use a definition introduced in \cite{RGflow} (see also \cite{chapterJan} for discussion), where "constant physics" relies on two quite natural assumptions, namely that (i) the  "shape" of the (average)  CDT  geometry and (ii) the (relative) amplitude of quantum fluctuations   remain constant on each RG flow trajectory. As will be shown below  such a definition has important consequences for the renormalized gravitational coupling constant if the semiclassical description in terms of the (Euclidean) de Sitter universe and the minisuperspace effective action  introduced in Section \ref{sec:2} stays valid up to  the  UV scale.

As discussed in Section \ref{sec:2}, the (average) CDT  geometry inside the semiclassical phase $C_{dS}$  seems to be consistent (at least when the scale factor is considered) with the (Euclidean) de Sitter universe, i.e., a regular four-sphere $S^4$. It implies a relation  between the measured parameter $\tilde \omega$, quantifying the width of the CDT geometry in lattice units, and its continuum counterpart $ \omega $, set by a condition that   the total four-volume of the sphere equals $V_4$, see equation \eqref{identification2},
$$
\tilde \omega = \omega C_4^{1/4}.
$$
There is also  a relation  between $\tilde\omega, \omega$ and the coupling constants $\tilde \mu, \mu$ appearing in the minisuperspace \eqref{CSMSV} and the CDT \eqref{eq:Sdiscrete}   effective actions, respectively, see equation \eqref{identification3a}, 
$$
  \tilde \mu\,\tilde  \omega^{8/3}  =    \mu\,   \omega^{8/3} = \text{const}.
$$
Following \cite{RGflow}, let us now consider anisotropic space-time scaling scenarios \`a la Ho\v{r}ava-Lifshitz gravity \cite{CDTHL, HLoriginal, chapterHL}.\footnote{One should note that such an approach manifestly breaks the four-dimensional diffeomorphism invariance of the continuum theory by introducing a preferred time-foliation and thus makes the asymmetry between space and time a real  physical phenomenon.} In the cosmological minisuperspace truncation one recovers the continuum semiclassical volume profile \eqref{CsolMSV}, as well as the continuum minisuperspace effective action \eqref{CSMSV}, but now $\omega$ becomes a free parameter, which  quantifies the space-time anisotropy, and consequently  the volume profile \eqref{CsolMSV} corresponds to a  deformed $S^4$ geometry, i.e., a four-ellipsoid, compressed or elongated in time direction depending on $\omega$ (for  $\omega = (\frac{3}{8 \pi^2})^{1/4}$ one recovers the symmetric  $S^4$). In such a case also the above relation \eqref{identification3a} between $\omega$ and $\mu$ stays valid, and thus the space-time asymmetry, quantified by $\omega$, is set by the value of (dimensionless) coupling constant $\mu$ appearing in the minisuperspace action \eqref{CSMSV} ($\mu = 9 (2 \pi^2)^{2/3}$ corresponds to a symmetric case consistent with General Relativity). As a result, though relation \eqref{identification2}, 
different  values of the lattice parameter $ \tilde \omega$  will  translate into  different values of the continuum parameter $\omega$ and thus to a physically different geometry of the quantum universe. Consequently   
\beql{ren4}
    \tilde \omega(k_0, \Delta) = \text{const}.
\eeq
is required to keep the (average) continuum geometry fixed and therefore it 
defines the lines of "constant physics" in the $(k_0, \Delta)$ coupling constant space.

In order to define the RG flow  one should also require that the relative amplitude of quantum fluctuations stays constant on each line of "constant physics" when approaching the continuum limit, i.e., when the lattice volume $N_4$ is increased and the lattice spacing $a$ is decreased in accordance with equation \eqref{ren1}.  The condition
\beql{ren55}
     \frac{\sqrt{\langle \delta V_3^{\, 2}(t_k) \rangle}}{\langle V_3(t_k) \rangle}=\frac{\sqrt{\langle \delta N^{\, 2}_3(k) \rangle_{\bar N_4}}}{\langle N_3(k)\rangle_{\bar  N_4} }=\text{const.} \quad , \quad t_k = k \, a
\eeq
 means that one investigates the same "real" continuum physics when changing the cutoff scale, and the result does not depend on $N_4$ nor $a$. 
 It cannot be achieved for fixed $k_0$ and $\Delta$, as in that case relative fluctuations vanish when $N_4\to \infty$, see equation  \eqref{eq:dV}. Accordingly, the RG flow towards a continuum limit requires  following a path of "constant physics" (defined by the condition \eqref{ren4}) in the CDT coupling constant space $(k_0(N_4),\Delta(N_4))$   when $N_4$ (and therefore $a\propto N_4^{-1/4}$) is changed.   The amplitude of spatial volume fluctuations is set by the (renormalized) dimensionless gravitational coupling constant $\tilde \Gamma$ (cf. the CDT effective action \eqref{eq:Sdiscrete}) and it scales as $\delta N_3\propto \sqrt{\tilde \Gamma} \, N_4^{1/2} $  (see equation \eqref{eq:Corrscale}). At the same time $N_3 \propto \tilde \omega^{-1} N_4^{3/4}  $ (see equation \eqref{eq:deSitter}). Therefore the condition \eqref{ren55}  implies 
\beql{ren3}
    \frac{ {\tilde \Gamma}\ \tilde\omega^2 }{N_4^{1/2}}=\text{const.}
\eeq
It means that when the lattice volume $N_4$ is changed one should follow a RG flow trajectory in the direction where  the (renormalized) {\it dimensionless} gravitational coupling constant $\tilde \Gamma$   scales as
\beql{ren66}
{\tilde \Gamma\big(k_0(N_4), \Delta(N_4)\big)} \propto N_4^{1/2},
\eeq
where we have explicitly included the dependence of $\tilde \Gamma$ on  $k_0$ and $\Delta$.
As the amplitude of fluctuations scales as $\delta N_3\propto \sqrt{\tilde \Gamma(k_0, \Delta)}$, the above condition compensates the $N_4^{-1/2}$ factor in equation \eqref{ren3}, resulting from a different scaling of the fluctuations and the average volume profiles with $N_4$. One should note that combining all RG flow conditions, i.e., equations  \eqref{ren1}, \eqref{ren4} and \eqref{ren66}, leads to
(cf. equation \eqref{eq:GGamma})
\beql{ren2}
    G\propto {\tilde \Gamma}\,  \tilde\omega^2 \, a^2  \propto \frac{ {\tilde \Gamma}\,  \tilde\omega^2} {N_4^{1/2}} = \text{const.}
\eeq
Therefore keeping the same continuum physics when approaching a continuum limit translates into keeping the (renormalized) {\it dimensionful} gravitational constant $G$ fixed, at least if our effective semiclassical description holds up to this limit.

Summarizing,  we  assume that  each RG flow trajectory is defined by a path in the CDT bare dimensionless  coupling constants space $(\kappa_0(N_4), \Delta(N_4))$ parametrized by $N_4$ (the lattice cutoff scale is $\propto N_4^{1/4}$) which fulfills  the following  conditions
\beql{ren5}
    \tilde \omega\big(k_0(N_4), \Delta(N_4)\big) = \text{const}.\quad , \quad  {\tilde \Gamma\big(k_0(N_4), \Delta(N_4)\big)} \propto N_4^{1/2}.
\eeq
If the path continues to $N_4\to \infty$ ($a\to 0$), which is possible at a higher-order critical  point  $(k_0^*,\Delta^*)$, then such a point will be a UV fixed point.

\begin{figure}[ht]
    \centering
      \includegraphics[width = 0.45\textwidth]{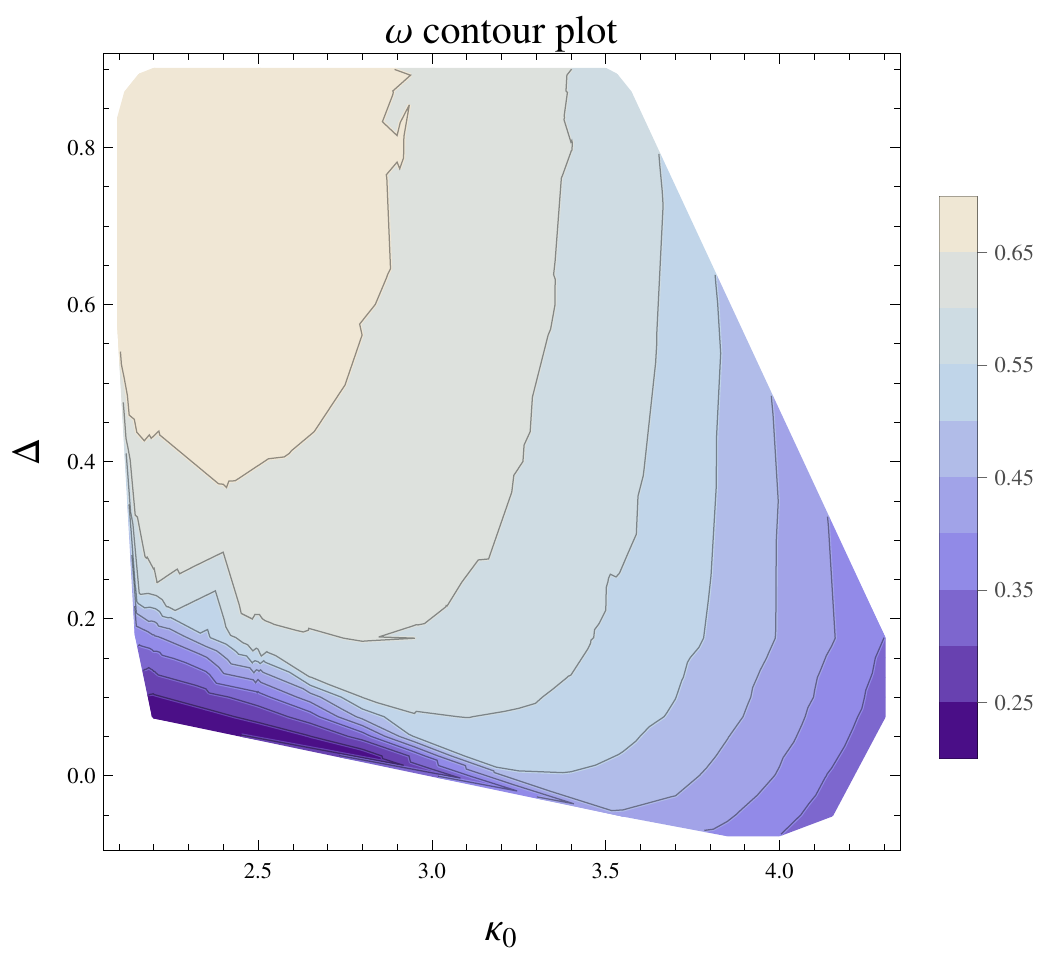}
      \includegraphics[width = 0.45\textwidth]{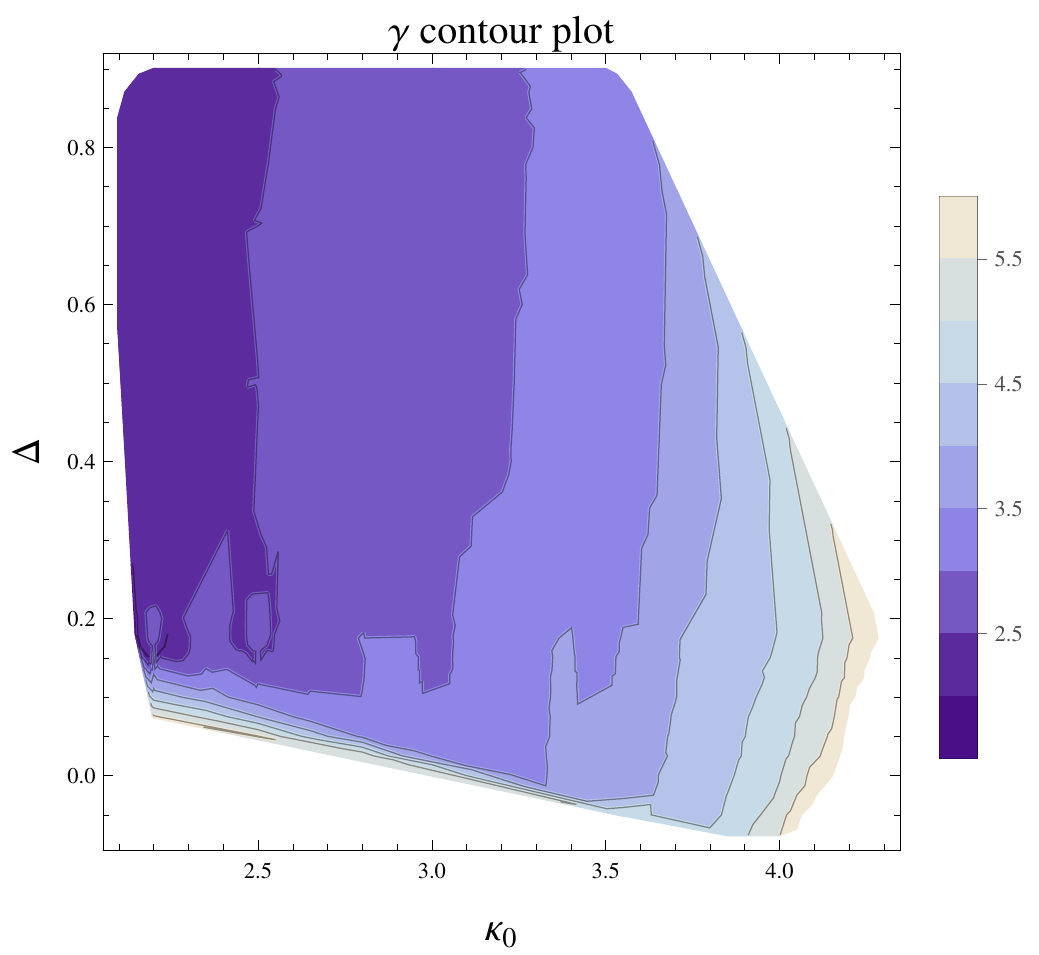}
    \caption{The "shape" parameter $\tilde \omega$ (left panel) and the fluctuations amplitude $ \gamma \propto \sqrt{\tilde \Gamma}$ (right panel) measured in the spherical CDT for $\bar N_4^{(4,1)}=40 \text{k}$. Courtesy of J. Ambj\o rn and A. G\"orlich.}
    \label{fig:RGflow}
\end{figure}

\begin{figure}[ht]    
    \centering
      \includegraphics[width = 0.6\textwidth]{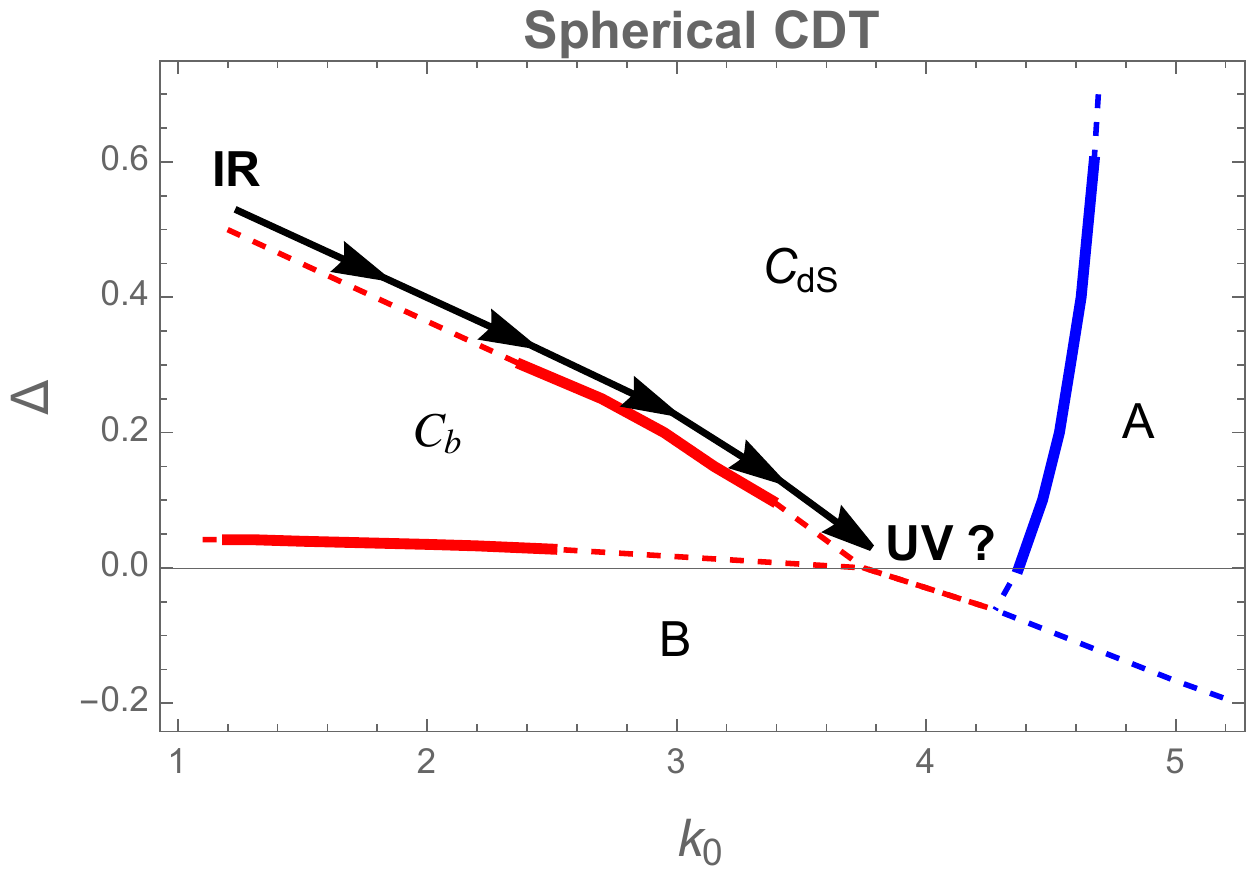}
    \caption{Putative CDT renormalization group flow line(s) leading from the IR to the (perspective) UV fixed point.}
    \label{fig:RGflow2}
\end{figure}

In order to check if the conditions  \eqref{ren5} can be satisfied one can perform a series of MC simulations in a grid of points $(k_0, \Delta)$ to  measure the distributions $N_3(k)$    and then compute $\tilde \omega (k_0, \Delta)$ and $\tilde \Gamma(k_0, \Delta)$. Our analysis assumes that $\tilde \omega (k_0, \Delta)$ and $\tilde \Gamma(k_0, \Delta)$ do not depend on $\bar N_4$ which, as  discussed in Section \ref{sec:2}, is indeed the case inside the phase $C_{dS}$ for large enough $\bar N_4$. It is therefore 
 enough to carry out measurements
for a fixed lattice volume $\bar N_4$. The results of such measurements \cite{RGflow} performed for $\bar N_4^{(4,1)}=40\text{k}$ are visualized in Fig. \ref{fig:RGflow}, where we show contour plots of $\tilde \omega (k_0, \Delta)$ (left panel) and
 $ \sqrt{\tilde \Gamma (k_0, \Delta)} $ (right panel) in the $(k_0, \Delta)$-plane. The lines visible on the left   plot can be directly interpreted as lines  of "constant physics", where $\tilde \omega (k_0, \Delta)=\text{const}$. 
 Moving along any such line one can read off from the right plot how $\sqrt{\tilde \Gamma (k_0, \Delta)}$ changes  and thus in which  direction one should follow  when  $N_4$ is increased (i.e., the lattice spacing $a$ is decreased). It is seen that the lattice spacing decreases towards the right-bottom corner of the plots, i.e., towards the interesting region of the CDT phase diagram, where  higher-order transition lines meet. 
 It is also seen that most of the RG flow trajectories, which are deep inside the $C_{dS}$ phase, go parallel to the $C_b-C_{dS}$ phase transition line but then turn away   
 and go parallel to the $A-C_{dS}$ phase  transition line.  Therefore such RG flow lines  do not lead to any UV fixed point. 
 However, there are potentially some lines of "constant physics" (very close to the $C_b-C_{dS}$ phase transition line) which may lead to a common critical point,  cf. Fig.~\ref{fig:RGflow2}.

\subsection{Summary of the search for a continuum limit}

Examples of simple lattice models, such as the Ising model, show that determining the order of a phase transition may be a difficult task and one has to carefully analyze finite-size effects. In the case of CDT  additional complication comes from the fact that the CDT phase transitions are not typical phase transitions studied in "standard" lattice QFTs, defined on the fixed spacetime, but they rather involve changes in  spacetime geometry or even  spacetime (effective) topology.
Nevertheless, using numerical MC simulations, one can study  properties  of the CDT phase transitions and in principle distinguish between the first- and the higher-order  transitions.
The results of the CDT phase transition studies performed so far lead to the conclusion that only in the case of the spherical CDT one can observe the higher-order phase transition lines (the $C_b-C_{dS}$ phase transition  and potentially the $B-C_{dS}$ phase transition) bordering the physically interesting   phase $C_{dS}$, inside which one observes the IR semiclassical universe. In that case one can also define the RG flow trajectories, related to keeping the physical "shape" of the universe and (relative) amplitude of quantum fluctuations fixed and leading in the direction of  decreasing lattice spacing. Combining RG flow conditions with an assumption that the physical four-volume of the universe is kept fixed  leads to a condition that the {\it dimensionful} gravitational coupling constant $G$ is  kept fixed on each RG flow line up to the UV scale, see equation \eqref{ren2}. One should note that this result is not compatible with the standard functional renormalization group approach to asymptotic safety, where the {\it dimensionless}  gravitational coupling stays non-zero but the dimensionful $G\to 0$ as the UV cutoff is removed. Thus comparing  results of both approaches is difficult and requires further studies.
Nevertheless, despite most of the RG flow  trajectories in CDT do not end up in a common  UV fixed point, there are some trajectories which potentially lead to a continuum limit, as  schematically outlined in Fig. \ref{fig:RGflow2}.
The  results are not conclusive as the precision of the MC measurements discussed above was not good enough to make sure that the RG flow really goes to the genuine UV fixed point(s).

\section{Conclusions}

Causal Dynamical Triangulations  is based on  lattice QFT techniques applied to the quantization of gravity and uses only a few additional assumptions about the topology of spacetime. 
Lattice QFTs turned out to be very successful in addressing many non-perturbative problems in other areas of physics, e.g., regarding Quantum Chromodynamics (QCD)  beyond perturbation regime. QCD is an ordinary QFT in flat spacetime and the regular hyper-cubic lattice used represents a simple discretization of the flat spacetime. However, if the field theory is gravity, spacetime itself becomes non-trivial and dynamical. Therefore
CDT generalizes the conventional  lattice QFT to the case of fluctuating geometries, approximated by lattices constructed from a few types of simplices with fixed edge lengths. Using Regge 
prescription of how to compute spacetime curvature for a simplicial manifold one can express  Einstein-Hilbert action in terms of dimensionless lattice variables.   The Regge action is defined  in terms of geometric invariants, such as lengths and angles, and does not use spacetime coordinates, thus making CDT    diffeomorphism-invariant.
CDT is also manifestly background independent as all possible (lattice regularized) spacetime geometries enter the quantum-gravitational path integral.
The key non-trivial assumption of CDT is that the quantum spacetime is globally hyperbolic and therefore can be foliated into spatial slices of constant cosmological time, which all have the same fixed topology.
Although the distinguished notion of  time in CDT looks superficially similar to the time-foliation in Horava-Lifshitz gravity \cite{HLoriginal}, its status is different because CDT does not possess any residual diffeomorphism invariance, which therefore cannot be broken either. 
The role of time in CDT was partly clarified in a study in three dimensions, where it was verified explicitly that key results of CDT quantum gravity continue to hold in a version of the theory which does not possess a preferred time foliation \cite{3dCDT}. This provides strong evidence that the notion of  time that is naturally available in standard CDT is simply a convenient foliation choice (a la gauge fixing) and that its presence does not affect the results of the theory in an unwanted way. The CDT time-foliation enables one to precisely define what is meant by a Wick rotation from the Lorentzian to the Euclidean spacetime signature and thus one can use the (Euclidean) model to investigate  properties of the quantum-gravitational path integral using numerical Monte-Carlo methods.

One of the greatest achievements  of CDT is that it correctly predicts the IR limit of quantum gravity, consistent with classical Einstein's field equations. The semiclassical limit is obtained in the, so-called, de Sitter phase $C_{dS}$, where one observes the emergent  four-dimensional average geometry with superimposed quantum fluctuations. 
What is more,  fluctuations of the global "shape" of the universe, quantified by the time evolution of the scale factor, are very well described by a simple minisuperspace effective action.
One should note that despite the fact that  the observed  semiclassical solution for the scale factor is consistent with a homogeneous and isotropic universe, the above mentioned symmetries are not put in by hand in CDT (as it is done in minisuperspace models), but emerge dynamically, as also does the effective spacetime dimension four. 
It is quite remarkable that despite the fact that the geometries entering the gravitational path integral are very  non-classical, after performing a suitable average over configurations  the scale factor gets consistent with a homogeneous and isotropic spacetime. 
However, since the scale factor is just one mode of the metric, in order to verify that the observed semiclassical ground state geometry is really homogeneous and isotropic one should use some other observables quantifying the "average geometry". This  is still an open issue, but some progress has been recently made in this direction, see \cite{chapterRenate, chapterWlosi, chapterATG}. Another non-trivial result is that the CDT effective action agrees with the standard Euclidean minisuperspace action  only up to overall sign (cf. equations \eqref{CSMSV} and \eqref{eq:Sdiscrete}, where both $\Gamma, \tilde \Gamma > 0$). 
The standard Euclidean minisuperspace action is unbounded from below due to negative sign of the kinetic term. The sign of the action does not matter for a classical trajectory but in the path integral formalism the unbounded action causes the path integral to be completely dominated by arbitrarily large fluctuations of the conformal mode,
 making the quantum theory ill-defined. A cure to the conformal mode problem can possibly come from  a specific contour of integration for the conformal factor  as Hartle and Hawking proposed in their minisuperspace truncation of the gravitational path integral \cite{HartleHawkingWF}. In CDT the problem is cured in a natural way without resorting to the Hartle-Hawking "trick", as one automatically obtains the inverted sign of the kinetic term,  exactly consistent with the Hartle-Hawking proposal. This simply comes from a  subtle interplay between the purely geometric degrees of freedom (entering the bare Regge action of CDT) and the entropic factor (counting the number of states with the same value of the bare action), 
which seems to stabilize the underlying quantum theory.

Another good feature of the four-dimensional CDT is that is has a reach phase structure, where (in the case of spherical CDT) some of the phase transitions bordering the physically interesting semiclassical phase $C_{dS}$ seem to be higher-order transitions. As a result one may hope to test the asymptotic safety scenario, where the UV fixed point of quantum gravity should appear as a higher-order critical point. 
Accordingly, one can try to define the RG flow of the CDT bare coupling constants towards the fixed point. 
If the RG flow leads to a higher-order critical point in such a way that the transition point could be associated with a gravitational UV fixed point one would obtain a non-perturbative description of a theory which can be viewed as the
quantum theory of gravity defined at arbitrarily short distances or, alternatively, at arbitrarily large energy scales. 
The results reported in Section \ref{sec:RGflow} are not conclusive as the resolution of numerical data is not good enough to make sure if (at least some) of the RG flow lines lead to a higher-order critical point.  
One should also note that the above results were obtained  \cite{RGflow} before the $C_b-C_{dS}$ phase transition was  discovered \cite{EffactionJGS}, and therefore at that moment the results were missing our current understanding of the CDT phase structure. It would be then worth to repeat the RG flow measurements using current computer resources and improved algorithms of the MC simulations which would enable  to study much larger systems thus reducing finite-size effects and increasing accuracy of the numerical results.
Having said this, let us comment about a possible scenario that the  continuum limit cannot be found.  First of all it can mean that our definition of what is considered to be a line of "constant physics" (here defined as a constant "shape" of the semiclassical CDT universe) is wrong and one should use some other notions of "constant physics", 
which can be well motivated by nonclassical features of quantum geometry  found on Planckian scales. This is indeed possible as the CDT approach seems to be non-compatible with the standard renormalization group techniques used in asymptotic safety. It leads again to the problem of defining good observables quantifying such features. But even in the most pessimistic scenario, where the new observables wouldn't help, CDT could still be considered  as an effective QFT of gravity,  valid up to some finite energy scale of order of the Planck's scale.

\printbibliography



\end{document}